\documentclass[twocolumn]{aastex6}
\usepackage{amsmath}
\usepackage{CJK}
\usepackage{natbib,hyperref}

\newcommand{\rp}{R_{\rm p}}
\newcommand{\rH}{r_{\rm H}}
\newcommand{\rB}{r_{\rm B}}
\newcommand{\rs}{r_{\rm s}}
\newcommand{\xs}{x_{\rm s}}
\newcommand{\cs}{c_{\rm s}}
\newcommand{\Mp}{M_{\rm p}}
\newcommand{\Mstar}{M_{\ast}}
\newcommand{\Me}{M_{\oplus}}
\newcommand{\qt}{q_{\rm thermal}}

\newcommand{\eqnref}[1]{Equation \ref{#1}}
\newcommand{\secref}[1]{\S\ref{#1}}
\newcommand{\figref}[1]{Figure \ref{#1}}
\newcommand{\tabref}[1]{Table \ref{#1}}

\begin{document}
\begin{CJK*}{UTF8}{bsmi}

\title{Circumplanetary Disk Dynamics in the Isothermal and Adiabatic Limits}

\author{Jeffrey Fung (馮澤之)\altaffilmark{1,2,3}, Zhaohuan Zhu (朱照寰)\altaffilmark{4}, Eugene Chiang (蔣詒曾)\altaffilmark{2,5}}
\altaffiltext{1}{Institute for Advanced Study, 1 Einstein Drive, Princeton, NJ 08540}
\altaffiltext{2}{Department of Astronomy, University of California, Campbell Hall, Berkeley, CA 94720-3411}
\altaffiltext{3}{NASA Sagan Fellow}
\altaffiltext{4}{Department of Physics and Astronomy, University of Nevada, Las Vegas, 4505 South Maryland Parkway, Las Vegas, NV 89154}
\altaffiltext{5}{Department of Earth and Planetary Science, University of California Berkeley, Berkeley, CA 94720-4767}

\email{email: fung@ias.edu}

\begin{abstract}
Circumplanetary disks (CPDs) may be essential to the formation of planets,
regulating their spin and accretion evolution.
We perform a series of 3D hydrodynamics simulations in both the isothermal and adiabatic limits to systematically measure the rotation rates, sizes, and masses of CPDs as functions of 
$\qt$, the ratio of the planet mass to the disk thermal mass. Our $\qt$ ranges from 0.1 to 4; for our various disk temperatures, this corresponds to planet masses between 1 Earth mass and 4 Jupiter masses. Within this parameter space, we find that isothermal CPDs are disky and bound within $\sim$10\% of the planet's Bondi radius $\rB$, with the innermost  $\sim0.05\,\rB$ in full rotational support. Adiabatic CPDs are spherical (therefore not actually ``disks''), bound within $\sim0.2\,\rB$, and mainly pressure-supported with rotation rates scaling linearly with $\qt$;
extrapolation suggests full rotational support of adiabatic envelopes at $\sim10\,\qt$. 
Fast rotation and 3D super-sonic flow render isothermal CPDs significantly different in structure from --- and orders of magnitude less massive than --- their 1D isothermal hydrostatic counterparts. Inside a minimum-mass solar nebula, even a maximally cooled, isothermal CPD around a 10 Earth-mass core may have less than 1 Earth mass, suggesting that gas giant
formation may hinge on 
angular momentum transport processes in CPDs.
Our CPD sizes and masses appear consistent with the 
regular satellites orbiting solar system giants.
\end{abstract}



\section{Introduction}
\label{sec:intro}

Planets still embedded in their natal protoplanetary disks (PPDs) can
continue to grow by accreting circumplanetary gas.
Such gas generally rotates about the planet, but is not necessarily in full rotational support. For simplicity, in this paper we refer to bound circumplanetary material
as circumplanetary disks (CPDs) regardless of the degree of rotation---indeed, one of our primary goals will be to measure rotation rates. 
Two candidate CPDs have recently been observed in the PDS 70 system \citep{Keppler18,Wagner18,Christiaens19,Haffert19}.

Numerical work has revealed that the flow pattern around embedded planets can be strongly three-dimensional (3D).
Gas tends to flow vertically toward the planet from the poles, and is expelled radially near the midplane. Qualitatively, this pattern persists
whether the gas is isothermal \citep{Machida08,Tanigawa12,Fung15,Ormel15,Bethune19}, isentropic \citep{Fung17a}, or is modeled with more sophisticated thermodynamics 
\citep{DAngelo13,Szulagyi16,Szulagyi17,Cimerman17,Lambrechts17,Schulik19}.  
In simulations where the planet is 
modeled as a sink cell,
the equatorial outflow is reduced or even stopped, but the inflow is still primarily vertical \citep{Bate03,DAngelo03,Paardekooper08}.

Simulations appear to disagree, however, about the magnitude
of rotation.
Some simulations have found
that CPDs are rotationally supported \citep{Tanigawa12,Wang14}.
Others have reported slower or even unmeasurably small
rotation \citep{Ormel15,Fung15,Cimerman17,Kurokawa18,Bethune19}. 
Meanwhile, \citet{Szulagyi16} and \citet{Szulagyi17} found a dependence
of rotation on
the temperature of the planetary core.
These studies differ in many respects. Not only do they model different planet masses and use different equations of state, but their numerical parameters also differ in terms of spatial resolution and the smoothing length used to model the planet's gravitational potential. These differences make it 
difficult
to synthesize a coherent picture of CPD dynamics. A related
unresolved issue is the CPD mass.
Some suggest that mass is zero, as the entire CPD is unbound \citep{Ormel15,Cimerman17,Bethune19}, a situation referred to
as ``atmospheric recycling.'' Others disagree 
\citep{DAngelo13,Lambrechts17,Lambrechts19}.

In this paper, we seek answers to these basic question. We will determine the sizes, masses, and rotation rates of CPDs by systematically exploring the parameter space from Earth-mass to multi-Jupiter-mass planets, embedded in disks of varying temperatures. We will also assess numerical convergence across different resolutions and different hydrodynamics codes. To begin, we give a quick overview of scales.

\subsection{CPD length scales and the disk thermal mass}
\label{sec:scales}

\citet{Machida08} demonstrated that the local, shearing sheet model of a planet embedded in a Keplerian disk can be described by a set of nondimensional equations that is characterized by a single parameter --- namely, the ratio between the planet mass and the ``disk thermal mass,'' which we write as:
\begin{equation}
\qt = \frac{M_{\rm p}}{\Mstar ~H_{\rm p}^{3}} = \frac{q}{H_{\rm p}^{3}} \, ,
\end{equation}
where $\Mp$ is the planet's mass, $\Mstar$ is the star's mass, $q$ is the planet-to-star mass ratio, and $H_{\rm p}$ is the disk aspect ratio evaluated at the planet's position. Here, $\qt$ is related to the fundamental length scales in CPD dynamics, which include the Hill radius $\rH$, the Bondi radius $\rB$, the scale height of the background disk $h_{\rm p} = H_{\rm p} \rp$, and the half width of the co-orbital horseshoe region $x_{\rm s}$. The Hill and Bondi radii are:
\begin{equation}
\rH = \rp\left(\frac{q}{3}\right)^{1/3} \, ,
\label{eq:rH}
\end{equation}
and
\begin{equation}
\rB = \frac{GM_{\rm p}}{\cs^2} = \rp\frac{q}{H_{\rm p}^2} \, ,
\label{eq:rB}
\end{equation}
where $G$ is the gravitational constant, $\rp$ is the radial location of the planet, and $\cs$ is the sound speed of the gas. 
To provide a sense of scale, we can also write:
\begin{equation}
\qt = 0.7\, \left(\frac{\Mp}{10\, \Me}\right) \left(\frac{\Mstar}{M_{\odot}}\right)^{-1} \left(\frac{H_{\rm p}}{0.035}\right)^{-3}  \, ,
\end{equation}
\begin{equation}
\rH = 0.031\, {\rm au}\, \left(\frac{\rp}{\rm au}\right) \left(\frac{\Mp}{10\, \Me}\right)^{1/3} \left(\frac{\Mstar}{M_{\odot}}\right)^{-1/3} \, ,
\label{eq:rH_num}
\end{equation}
and
\begin{equation}
\rB = 0.027\, {\rm au}\, \left(\frac{\Mp}{10\, \Me}\right) \left(\frac{\cs}{\rm 1\, km/s}\right)^{-2} \, .
\label{eq:rB_num}
\end{equation}

For $x_{\rm s}$, \citet{Masset06} found that it can be separated into two regimes:
\begin{equation}
\xs = \begin{cases} \rp\sqrt{q/H_{\rm p}} &\mbox{if } \qt \lesssim 1 \, ,\\
2.5 \rH & \mbox{if } \qt \gtrsim 1 \, .\end{cases}
\label{eq:xs}
\end{equation}
The ratio between any two of these four length scales, $\rH$, $\rB$, $h_{\rm p}$, and $x_{\rm s}$, can be expressed in terms of $\qt$ and $\qt$ only. For instance, $\rB/\rH = 3^{1/3} \qt^{2/3}$, $h_{\rm p}/\rH = 3^{1/3} \qt^{-1/3}$, and $\xs/\rH = 3^{1/3} \qt^{1/6}$ in the $\qt\lesssim1$ regime. Hence, $\qt$ alone should be sufficient to determine the dynamics of the CPD, as long as we restrict ourselves to considering only gravity and hydrodynamics. While it is beyond the scope of this work, we note that other physical parameters such as thermal diffusivity and optical thickness will add extra dimensions to this problem. We focus here on how CPD dynamics depends on
$\qt$, and will verify that simulations with the same $\qt$ but different $q$ and $H_{\rm p}$ will yield the same results.

\section{Simulation Setup}
\label{sec:setup}

\subsection{Code description: \texttt{PEnGUIn}}
\label{sec:penguin}

We use the graphics processing unit-accelerated hydrodynamics code \texttt{PEnGUIn} \citep{MyThesis} to simulate planets embedded in disks. \texttt{PEnGUIn} solves the Euler equations:
\begin{align}
\label{eq:cont_eqn}
\frac{D\rho}{Dt} &= -\rho\left(\nabla\cdot\mathbf{v}\right) \,,\\
\label{eq:moment_eqn}
\frac{D\mathbf{v}}{Dt} &= -\frac{1}{\rho}\nabla p  - \nabla\Phi \,, \\
\label{eq:energy_eqn}
\frac{De}{Dt} &= -\frac{1}{\rho}\nabla\cdot p\mathbf{v}  - \nabla\Phi\cdot\mathbf{v} \,,
\end{align}
where $\rho$ is the gas density, $p$ is the gas pressure, $\mathbf{v}$ is the gas velocity, and $e = u + |\mathbf{v}|^{2}/2$ is the specific total (internal + kinetic) energy. In isothermal simulations, we discard \eqnref{eq:energy_eqn} and set $p/\rho$ to be globally constant. Here, $\Phi$ is the combined gravitational potential of the star and the planet. In spherical coordinates centered on the star, where \{$R$, $\psi$, $\theta$\} represent the radial, azimuthal, and polar coordinates respectively, $\Phi$ can be written as: 
\begin{align}
\nonumber
\Phi &= -\frac{G\Mstar}{R} - \frac{G\Mp}{\sqrt{R^2 + \rp^2 - R \rp \sin{\theta} \cos{\psi'} + \rs^2}} \\
\label{eq:grav}
~ &+ \frac{G\Mp R \sin{\theta}\cos{\psi'}}{\rp^2} \, ,
\end{align}
where $\psi' = \psi-\psi_{\rm p}$ is the angular distance from the planet. The third term on the right is the indirect potential due to our frame centering on the star rather than the center of mass. We have assumed that the planet is on a circular orbit in the midplane. The smoothing length $\rs$ is included to prevent numerical instability very close to the planet, 
and could represent the size of the planet's solid core. For example, the radius of the Earth at 1 au is about 1.6\% of the Earth's Bondi radius. In this work, we do not assume a size for the core, and instead aim to quantify how our results depend on $\rs$.
Following \citet{Fung17a}, we set $\rs$ to be the length of 3 grid cells (resolution is described in \secref{sec:resolution}; also see \tabref{tab:para} for values of $\rs$). 

Simulations are performed in a spherical grid centered on the star, and in a rotating frame at the planet's orbital frequency $\Omega_{\rm p}$, fixing the planet in space. This is advantageous because it reduces numerical diffusion caused by advection. The Coriolis force due to frame rotation is not explicitly computed; rather, it is absorbed into the conservative form of the angular momentum equation \citep{Kley98}.

For the equation of state (EOS), we use the ideal gas law, such that:
\begin{equation}
p = c_{\rm iso}^2 \rho \,,   
\end{equation}
in isothermal simulations, where $c_{\rm iso}$ is the isothermal sound speed, and 
\begin{equation}
p = u\rho(\gamma-1) \,,
\end{equation}
in adiabatic simulations, where $\gamma$ is the ratio of specific heats (or adiabatic index). We choose $\gamma=7/5$.

Mass and momentum fluxes across simulation cells are conservative, but the total energy is not. In PPDs, the orbital speed is highly supersonic, making the kinetic energy dominant over the internal energy by orders of magnitude. Thus, a conservative scheme in total energy would produce significant noise in the internal energy of the gas. We therefore opt to conserve the internal energy instead. Testing shows that this leads to significantly more stable flows.

In most cases, we run our simulations to 11 or 21 planetary orbits ($t_{\rm sim}$ in \tabref{tab:para}). In fact, they all appear to reach quasi-steady states even after just $\sim2$ orbits, in the sense that our results would not be significantly different even if we had terminated them at just $2$ orbits. Nonetheless, we run them much longer to confirm our results are robust. There are noticeable temporal fluctuations, particularly in the isothermal cases. These fluctuations are not qualitatively important, but can affect our quantitative results. Therefore, unless otherwise stated, all our results are time-averaged over the last orbit.

Even though our simulation grid centers on the star, for analysis it is convenient to use a cylindrical coordinate centered on the planet. Thus, we will also be using \{$r$, $\phi$, $z$\} to denote the radial, azimuthal, and vertical coordinates where the planet is at $r=z=0$.

\begin{deluxetable*}{cccccccc}
\tablecaption{\label{tab:para} Model Parameters}
\tablehead{Model \#& $H_{\rm p}$ & EOS & $\Mp$ & $\qt$ & Resolution & $\rs$\tablenotemark{a} & $t_{\rm sim}$\tablenotemark{b}\\
 & & & ($\Me$) &  & (cells/${\rm min}[\rB, h]$) & ($\rB$) & ($2\pi\Omega_{\rm p}^{-1}$)}
\startdata 
1 & 0.035 & isothermal & 1.4 & 0.1  & 64  & 0.047 & 21\\
2 & 0.035 & isothermal & 3.5 & 0.25 & 64  & 0.047 & 21\\
3 & 0.035 & isothermal & 7   & 0.5  & 64  & 0.047 & 21\\
4 & 0.035 & isothermal & 14  & 1    & 64  & 0.047 & 21\\
5 & 0.035 & isothermal & 28  & 2    & 64  & 0.023 & 21\\
6 & 0.035 & isothermal & 56  & 4    & 64  & 0.012 & 21\\
\hline
7  & 0.035 & adiabatic & 1.4 & 0.1  & 64  & 0.047 & 11\\
8  & 0.035 & adiabatic & 3.5 & 0.25 & 64  & 0.047 & 11\\
9  & 0.035 & adiabatic & 7   & 0.5  & 64  & 0.047 & 11\\
10 & 0.035 & adiabatic & 14  & 1    & 64  & 0.047 & 11\\
11 & 0.035 & adiabatic & 28  & 2    & 64  & 0.023 & 11\\
12 & 0.035 & adiabatic & 56  & 4    & 64  & 0.012 & 11\\
\hline
13 & 0.1 & isothermal & 333  & 1 & 64  & 0.023 & 21\\
14 & 0.1 & isothermal & 1333 & 4 & 64  & 0.012 & 100\tablenotemark{c}\\
\hline
15 & 0.1 & adiabatic & 333  & 1 & 64  & 0.023 & 11\\
16 & 0.1 & adiabatic & 1333 & 4 & 64  & 0.012 & 11\\
\hline
17\tablenotemark{d} & 0.035 & isothermal & 1.4 & 0.1  & 512\tablenotemark{e} & 0.006 & 3\tablenotemark{f}\\
18 & 0.035 & isothermal & 14  & 1    & 512\tablenotemark{e} & 0.006 & 3\tablenotemark{f}\\
\enddata
\tablenotetext{a}{The smoothing length $\rs$ is equal to 3 times the size of the smallest cells.}
\tablenotetext{b}{Unless otherwise stated, results are time-averaged over the last orbit.}
\tablenotetext{c}{Results are time-averaged between the 20\textsuperscript{th} to 21\textsuperscript{st} orbit, the same as other isothermal runs, but we then extend it to 100 orbits to study the effects of gap-opening (see \secref{sec:gap}).}
\tablenotetext{d}{Model \#17 is also simulated with \texttt{Athena++} using the same physical parameters but a different numerical setup. See \secref{sec:athena++}.}
\tablenotetext{e}{Unlike other models where the resolution is nearly uniform inside all of $\rB$, in models \#17 and \#18, only within $\sim0.1\,\rB$ is the resolution equivalent to 512 cells$/\rB$.}
\tablenotetext{f}{In models \#17 and \#18, we do not perform time averaging over the last orbit.}
\end{deluxetable*}

\subsubsection{Initial and boundary conditions}
\label{sec:condition}

We assume the initial disk is axisymmetric, is in hydrostatic equilibrium with the star's gravity, and has a power-law profile in the radial direction:
\begin{equation}
\label{eq:init_den}
\rho = \rho_0 \left(\frac{R \sin{\theta}}{\rp}\right)^{-3} \exp\left[-\frac{G\Mstar}{R c_{\rm iso}^2}\left(\frac{1}{\sin{\theta}} - 1\right)\right] \, ,
\end{equation}
where $\rho_0$ is a background normalization
(i.e., the ambient disk density at the planet's orbital radius,
not including perturbations by the planet). In the code, $\rho_0$
is set to 1; the exact value is immaterial because gas self-gravity
is neglected.

The power of $-3$ is chosen such that the gas surface (vertically integrated) density $\Sigma$ scales as $R^{-3/2}$. 
We also denote $\Sigma_0$ as the surface density at the planet's location. 
We choose $c_{\rm iso}$ to be either $0.035 \rp\Omega_{\rm p}$ or $0.1 \rp\Omega_{\rm p}$ (see \tabref{tab:para}), 
which respectively correspond to a disk aspect ratios $H_{\rm p}$ of 0.035 or 0.1 at the planet's location. 
To make the comparison as direct as possible, \eqnref{eq:init_den} is used for both isothermal and adiabatic simulations, 
and gas pressure is also initialized as $p=c_{\rm iso}^{2}\rho$ in both cases. In other words, the initial conditions in both the isothermal and adiabatic simulations are completely identical.
We note that the sound speed for adiabatic gas is not $c_{\rm iso}$, but rather $\cs=\sqrt{\gamma p/\rho}=\gamma^{1/2}c_{\rm iso}$. To keep our notation simple, the Bondi radius is evaluated as $\rB = G\Mp/c_{\rm iso}^2$ regardless of the EOS.

To establish a hydrostatic disk, initially there is no radial or polar motion, and the azimuthal rotation frequency is:
\begin{equation}
\label{eq:init_omg}
\Omega = \sqrt{\frac{G\Mstar}{R^3} + \frac{1}{R\rho}\frac{\partial p}{\partial R}} \, .
\end{equation}

Our simulation domain spans $\rp-10\,\rB$ to $\rp+10\,\rB$ radially, the full $2\pi$ azimuthally, and $\pi/2-3\,H_{\rm p}$ to $\pi/2$ in the polar direction, which is from the disk midplane to about 3 scale heights above.

We impose periodic boundary conditions in the azimuthal direction, and reflective boundaries in the polar direction. Reflective boundaries are used in the midplane to enforce symmetry, and at the top to prevent gas from flowing in or out of the simulation box. The radial boundaries are fixed to the initial values. Additionally, we place wave-killing zones next to the radial boundaries to help reduce wave reflections. They are prescribed as follows:
\begin{equation}
\label{eq:kill}
\frac{\partial X}{\partial t} = \frac{X(t=0) - X}{t_{\rm kill}} \left(1-\frac{|r-r_{\rm bound}|}{L_{\rm kill}}\right)^2 \, ,
\end{equation}
where $X$ corresponds to the fluid properties $\rho$, $p$, and each component in $\mathbf{v}$; $t_{\rm kill}$ is the damping timescale; $r_{\rm bound}$ is the position of either the inner or the outer radial boundary; and $L_{\rm kill}>|r-r_{\rm bound}|$ is the width of the kill zone. We choose $t_{\rm kill} = 0.2\,\pi\,\Omega_{\rm p}^{-1}$ which is one-tenth of the planet's orbital period, and $L_{\rm kill}=h$, except when $\qt=0.1$ (models \#1, 7, and 17), where we have $L_{\rm kill}=0.1~h$ instead.

\subsubsection{Resolution}
\label{sec:resolution}

We use a nonuniform grid to concentrate resolution near the planet. If we denote $L$ as the maximum distance away from the planet along one of the three coordinates, $N$ as the number of cells within $L$, and $\rm i$ as the $\rm i^{\rm th}$ cell away from the planet, then $x_{\rm i}$, the distance from the $\rm i^{\rm th}$ cell to the planet, is:
\begin{equation}
\label{eq:res}
x_{\rm i} = {\rm i} \Delta x_{\rm min} + \left(L-N\Delta x_{\rm min}\right) \left(\frac{i}{N}\right)^{a} \,,
\end{equation}
where
\begin{equation}
a = \frac{\ln\left(1-\frac{\Delta x_{\rm max} - \Delta x_{\rm min}}{L-N\Delta x_{\rm min}}\right)}{\ln\left(1 - \frac{1}{N}\right)} \,.
\end{equation}
Here, $\Delta x_{\rm min}$ determines the resolution near the planet, while $\Delta x_{\rm max}\gg\Delta x_{\rm min}$ is the cell size farthest away from the planet. When $\qt\leq 1$ and $H_{\rm p}=0.035$, along each of the three directions \{$R$, $\psi$, $\theta$\}, we have $L=\{10\rB,~\pi,~3H_{\rm p}\}$, $\Delta x_{\rm max}=\{\rB/4,~0.1,~H_{\rm p}/8\}$, and $\Delta x_{\rm min}$ is either $\rB/64$ or $\rB/512$ (see \tabref{tab:para}) but is the same in all directions. When $H_{\rm p}=0.1$, $L$ in the radial direction is 0.6 $\rp$ instead. When $\qt=\rB/h>1$, we use $\Delta x_{\rm max}=\{h/4,~0.1,~H_{\rm p}/8\}$, and $\Delta x_{\rm min}=h/64$. We use $N=\{192,~240,~128\}$ in models \#1--16, and $N=\{224,~272,~192\}$ in models \#17--18; in terms of the total number of cells in the grid, they correspond to $384\times480\times128$ and $448\times544\times192$, respectively.

\begin{figure}
\includegraphics[width=0.99\columnwidth]{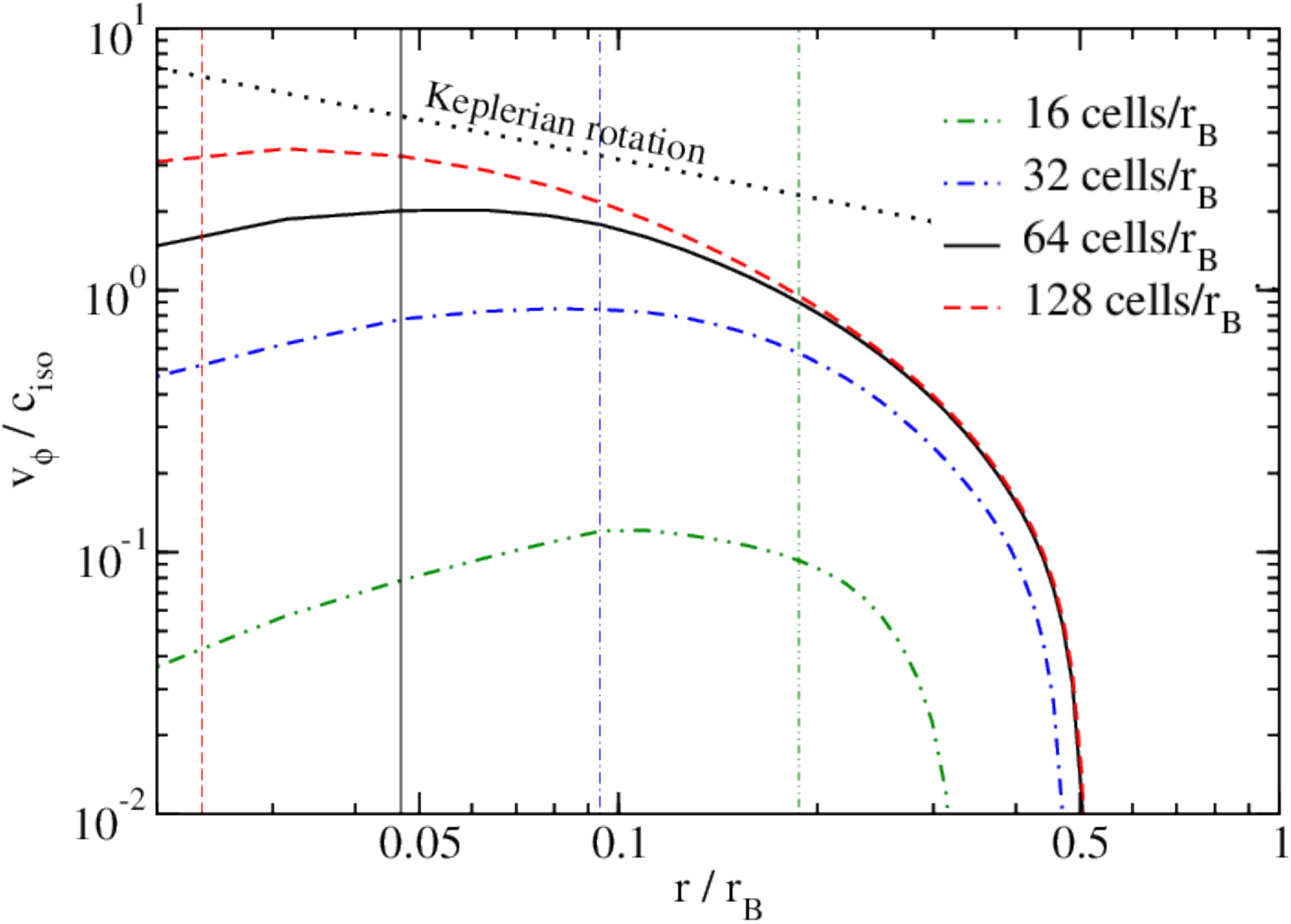}
\caption{Convergence with resolution for when $\qt=1$, $H_{\rm p}=0.035$, and with an isothermal EOS. Black curve (64 cells$/\rB$) comes from model \#4 and is our fiducial setup, which has reached convergence to a percent level for the bulk of the Bondi sphere. Black dotted line shows the Keplerian speed for reference. In all cases, the smoothing length $\rs$ for the planet's potential is equal to 3 times the cell size, and they are shown as vertical lines with the corresponding colors. Shorter $\rs$ (higher resolution) leads to faster rotation speeds close to the planet.}
\label{fig:res}
\end{figure}

A main goal of this work is to resolve and analyze the rotation in the gas around protoplanets. To confirm we can achieve this goal, we take our fiducial model where $\qt=1$ and $H_{\rm p}=0.035$ (the same as models \#4 and \#18), and simulate it under different resolutions ranging from 16 to 128 cells$/\rB$. The smoothing length $\rs$ is always set to be 3 times the cell size. \figref{fig:res} plots the rotation curves around the planet from these simulations. At 64 cells$/\rB$, we find that we have reached numerical convergence to within a percent level for the bulk of the Bondi sphere; while very close to the planet, 
$\lesssim 0.1\rB$, we start seeing the effects of the smoothed gravitational potential and find slower rotation speeds. Using this test as a guide, we use 64 cells$/\rB$ as our fiducial resolution (64 cells$/h$ when $\rB>h$, as in the cases with models \#5, 6, 11, 12, 14, and 16) to study dynamics on the $\rB$ scale, and enhance our resolutions by factors up to 8 in some models to study dynamics deep within $\rB$.

\subsection{Code description: \texttt{Athena++}}
\label{sec:athena++}

For one particular model, Model 17 in \tabref{tab:para}, we carry out a similar simulation
but with a quite different numerical setup using the \texttt{Athena++} code \citep{stone2008}. In contrast to the \texttt{PEnGUIn} setup, we adopt a spherical-polar grid centered on the planet. 
In the radial direction, the grid spacing is uniform in logarithmic space, with 256 cells from $r_{\rm min} = 3\times10^{-5}$ $\rp$ to $r_{\rm max}=0.35$ $\rp$. 
At the inner boundary, a reflecting boundary condition is adopted. At the outer boundary, the fluid variables are fixed to their initial values so that the gas flow is
orbiting around the central star.
In the polar direction, the grid is uniform with 32 grids from 0 to $\pi/4$. The polar boundary condition \citep{zhu2018} is adopted at the pole while a reflecting boundary condition is adopted at the midplane. In the azimuthal direction, the grid is uniform with 128 grids from 0 to 2$\pi$.
With this grid structure, each cell at the disk midplane has the same length in all three directions. At the inner boundary where resolution is the highest,
the edge of each cell has a length of $\rB/2378$.

The planet's gravitational acceleration is smoothed by the following function:
\begin{equation}
f=\frac{(r-r_{\rm min})^2}{(r-r_{\rm min})^2+r_{\rm smooth}^2}
\end{equation}
where $r_{\rm smooth}=4\times10^{-6}$ $R_{\rm p}$.
A density floor of $10^{-10}$ times the initial midplane density at $R_{p}$ is adopted.
All other planet and disk setups are the same as  the \texttt{PEnGUIn} setup as described in \secref{sec:condition}.

We run the simulation for 1.1 planetary orbits. If the CPD were rotating at Keplerian speed, it would correspond to 1.37$\times10^4$ orbits at $r_{\rm min}$. 

\section{Results}
\label{sec:result}

\begin{figure*}
\includegraphics[width=1.99\columnwidth]{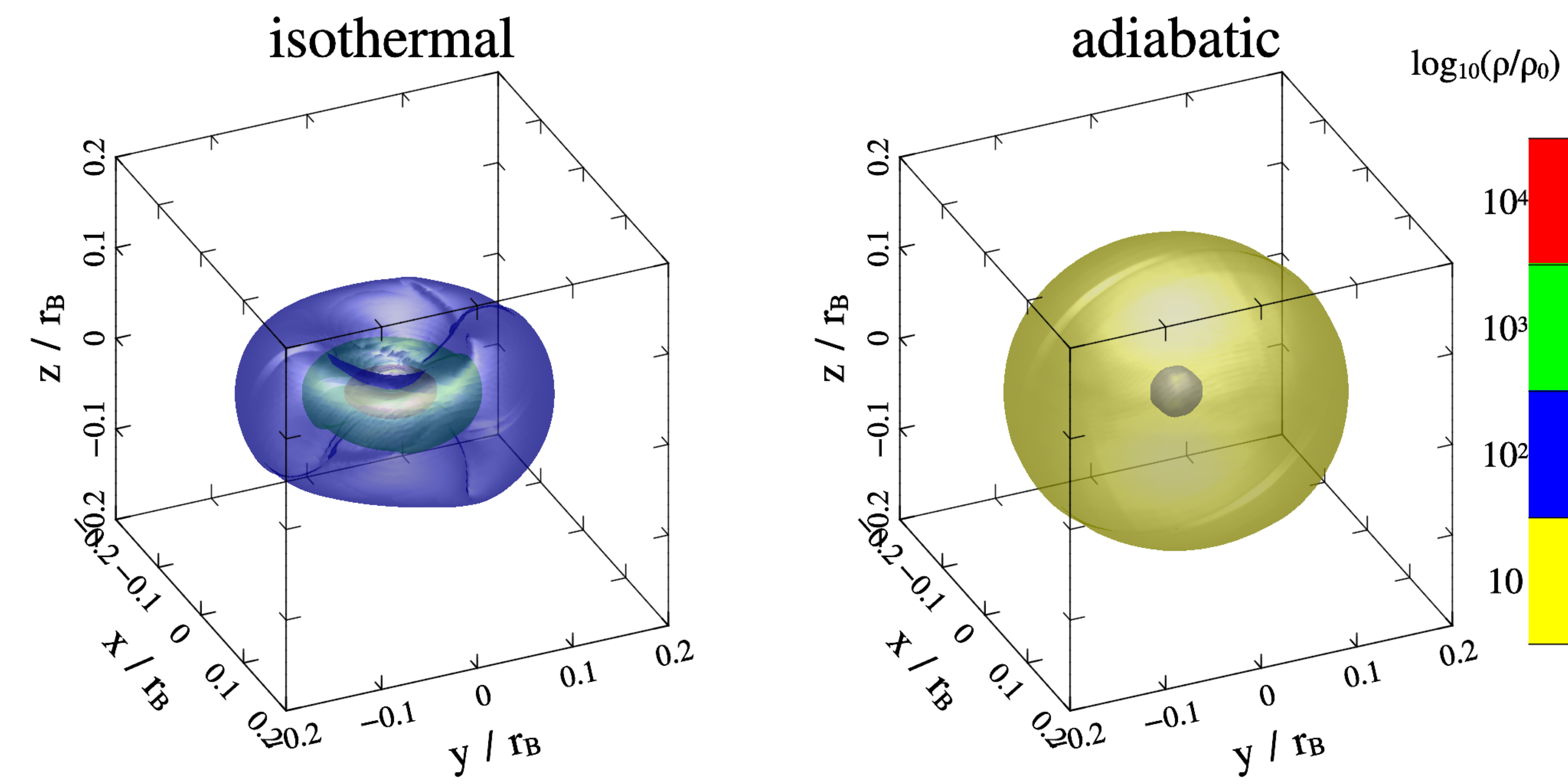}
\caption{Snapshots of 3D isodensity surfaces illustrating the typical morphologies of our CPDs. Isothermal CPDs are typically disk-like, whereas adiabatic CPDs are typically spherically symmetric. Left panel is taken from model \#18 at $t=3$ orbits, and right is from model \#10 at $t=10$ orbits. Both models have $\qt=1$. Yellow, blue, green, and red surfaces indicate $10$, $10^2$, $10^3$, and $10^4$ times the initial density $\rho_0$ at the planet's location without the planet's perturbation, respectively. The yellow surface is larger than the box of the plot in the left panel. The star is along the y-axis in the negative direction.}
\label{fig:3D_surfaces}
\end{figure*}

\begin{figure}
\includegraphics[width=0.99\columnwidth]{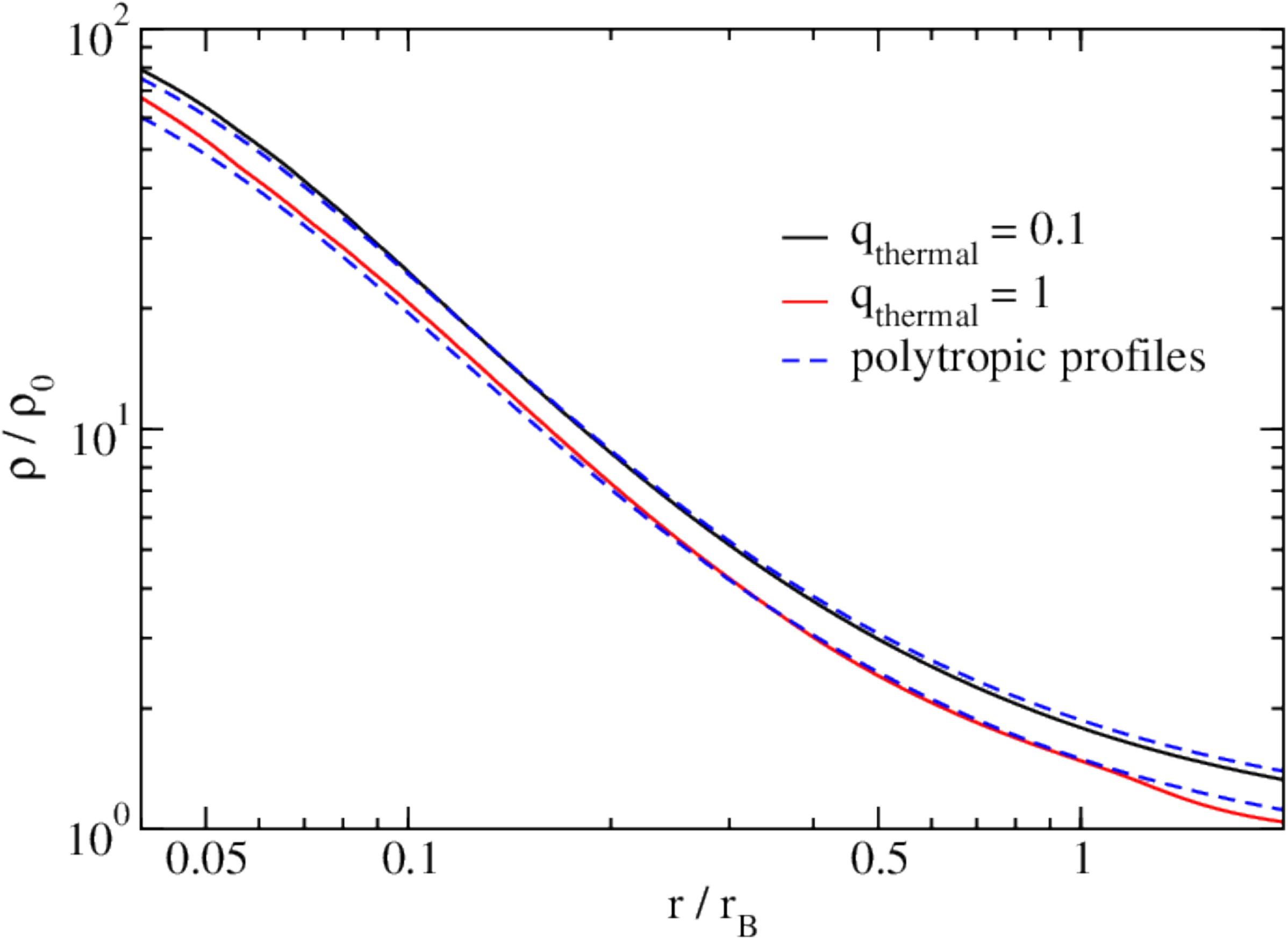}
\caption{Midplane density profile from adiabatic models \#7 (black) and \#10 (red), azimuthally averaged over $\phi$. Blue dashed curves are the hydrostatic polytropic profiles described by \eqnref{eq:poly_den}.
Both models follow the polytropic profile closely, although their normalizations are offset by about 20\%. This difference is due to planets of different $\qt$'s merging with the background PPD at different distances.}
\label{fig:ad_den_1D}
\end{figure}

\begin{figure}
\includegraphics[width=0.99\columnwidth]{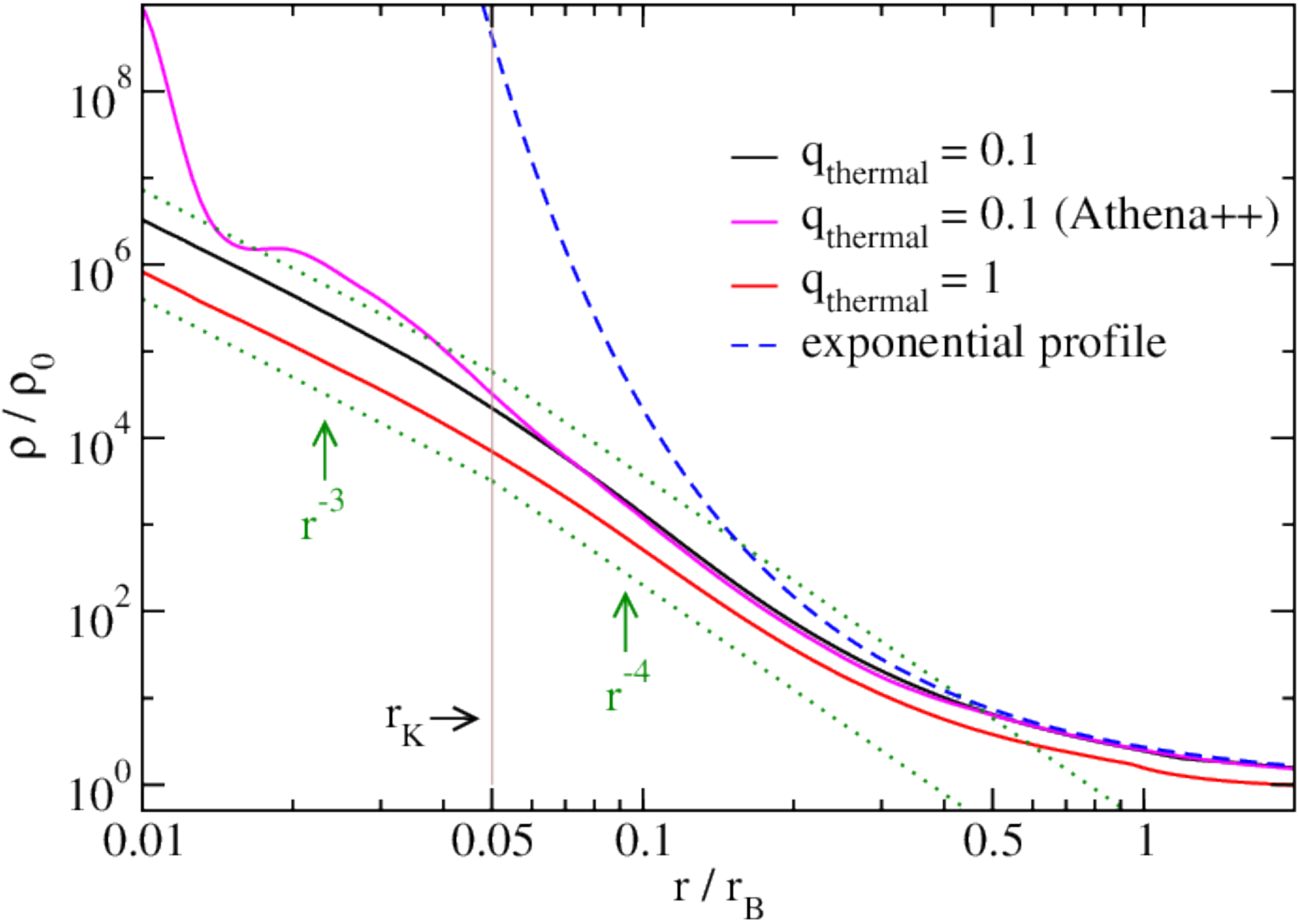}
\caption{Midplane density profile from
isothermal
models \#17 and \#18, azimuthally averaged over $\phi$. The blue dashed curve is the hydrostatic profile described by \eqnref{eq:iso_den}, which the actual density profiles do not follow. Instead, they can be roughly described as a broken power law that goes as $r^{-3}$ inside $r_{\rm K} = 0.05\rB$, and $r^{-4}$ between 0.05 to 0.2 $\rB$.
This CPD is rotationally supported within the radius $r_{\rm K}\sim0.05\,\rB$ (see \secref{sec:rotation_iso}).
The smoothing length $\rs$ is $\sim0.006\,\rB$ in these 2 models (\tabref{tab:para}), well within $r_{\rm K}$.}
\label{fig:iso_den_1D}
\end{figure}

\begin{figure}
\includegraphics[width=0.99\columnwidth]{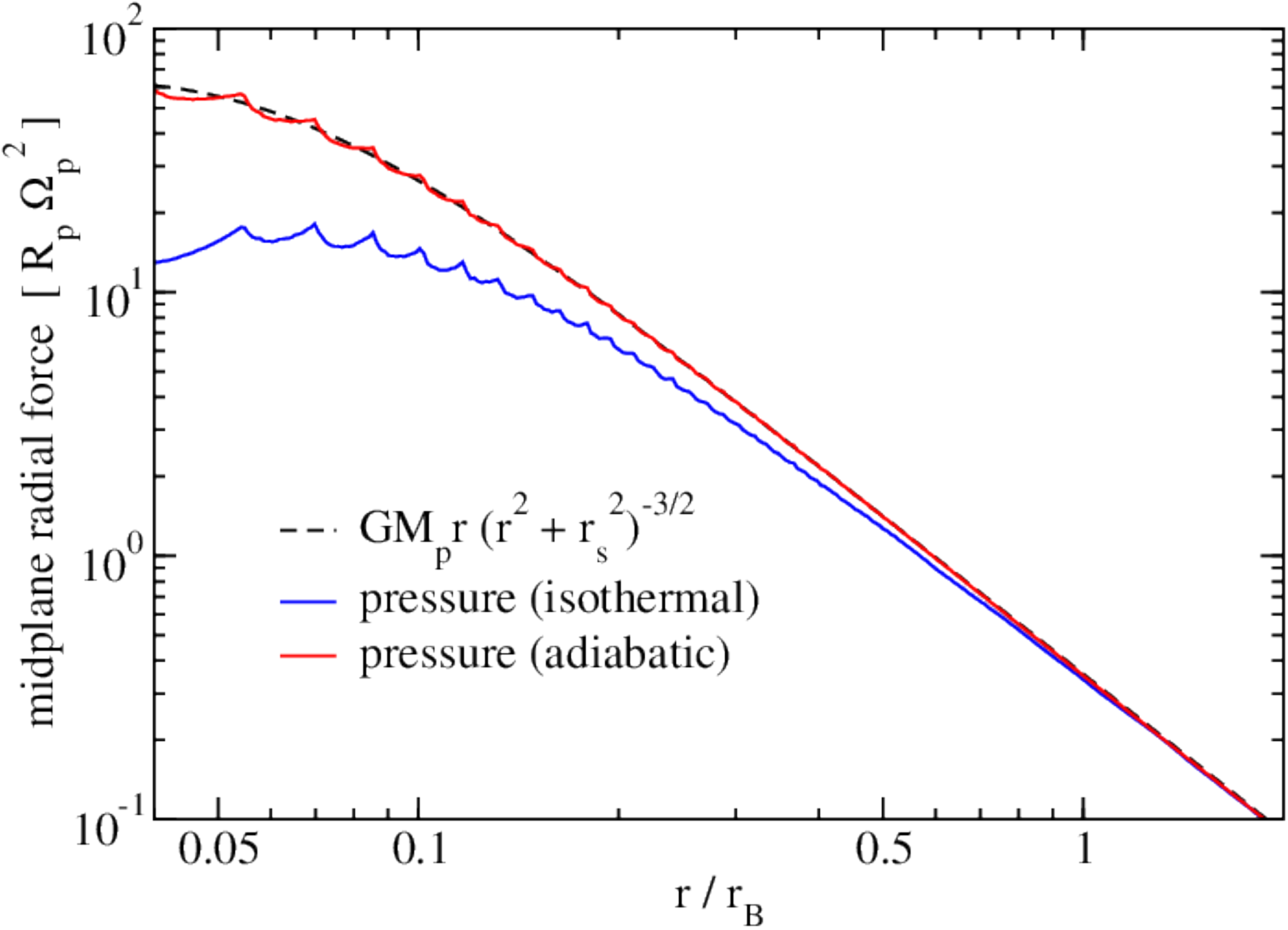}
\caption{Pressure support in the midplane, $-\frac{1}{\rho}\frac{\partial p}{\partial r}$, as a function of distance to the planet for cases where $\qt=0.1$. The profiles are azimuthally averaged over $\phi$. Blue is the isothermal simulation given by model \#1 and red is the adiabatic model \#7. Also shown is the gravitational force from the planet (dashed black). For an adiabatic CPD, the gas is supported by the radial pressure gradient; but when it is isothermal, however, pressure support weakens and we find that the gas is either rotationally supported or is in a steady flow.}
\label{fig:pressure}
\end{figure}

\begin{figure*}
\includegraphics[width=1.99\columnwidth]{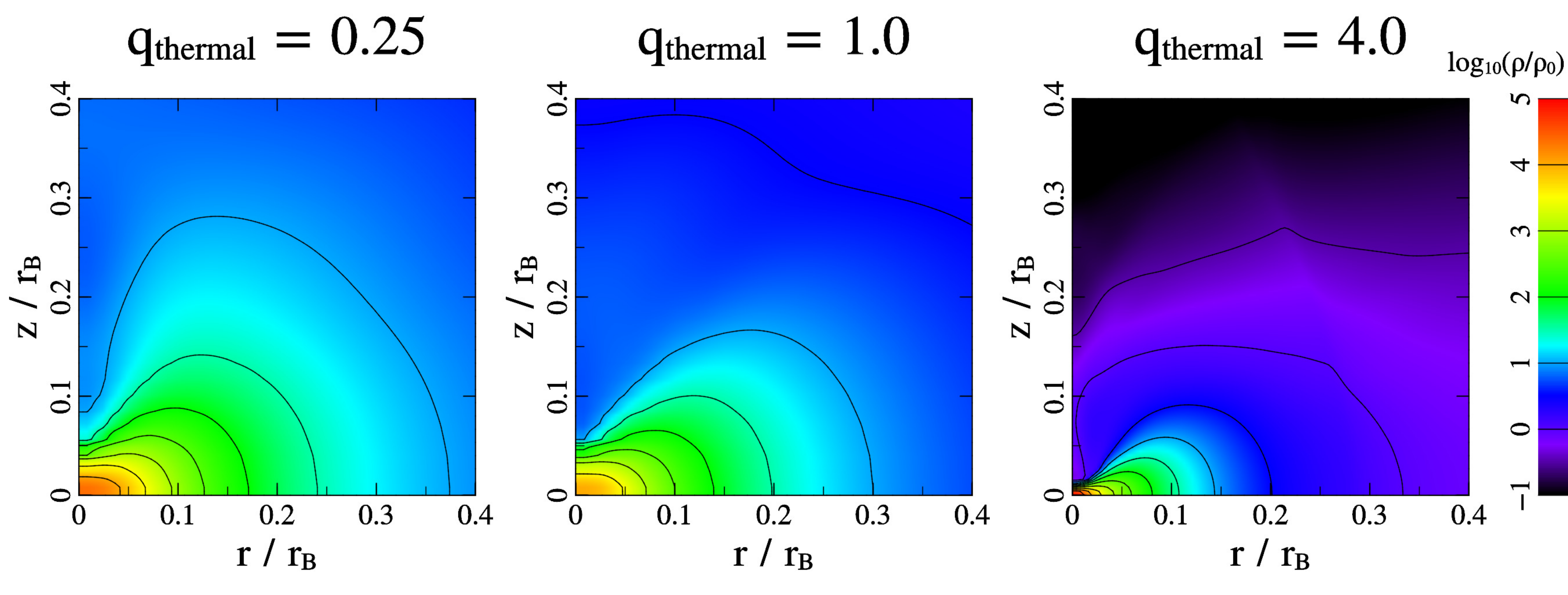}
\caption{Density from the isothermal simulations, azimuthally averaged over $\phi$. We plot results from models \#2 (left), \#4 (middle), and \#6 (right). Black lines are contours at intervals of 0.5 in logarithmic scale. Isothermal CPDs are disk-like, and the flattening is clear out to $\sim 0.4 \rB$. On the $\rB$ scale, the CPD size remains about constant when $\qt$ is subthermal, but shrinks as $\qt$ increases to superthermal values. This implies the CPD size may be scaling with $\rH$ instead in this regime.}
\label{fig:iso_den_2D}
\end{figure*}

\begin{figure*}
\includegraphics[width=1.99\columnwidth]{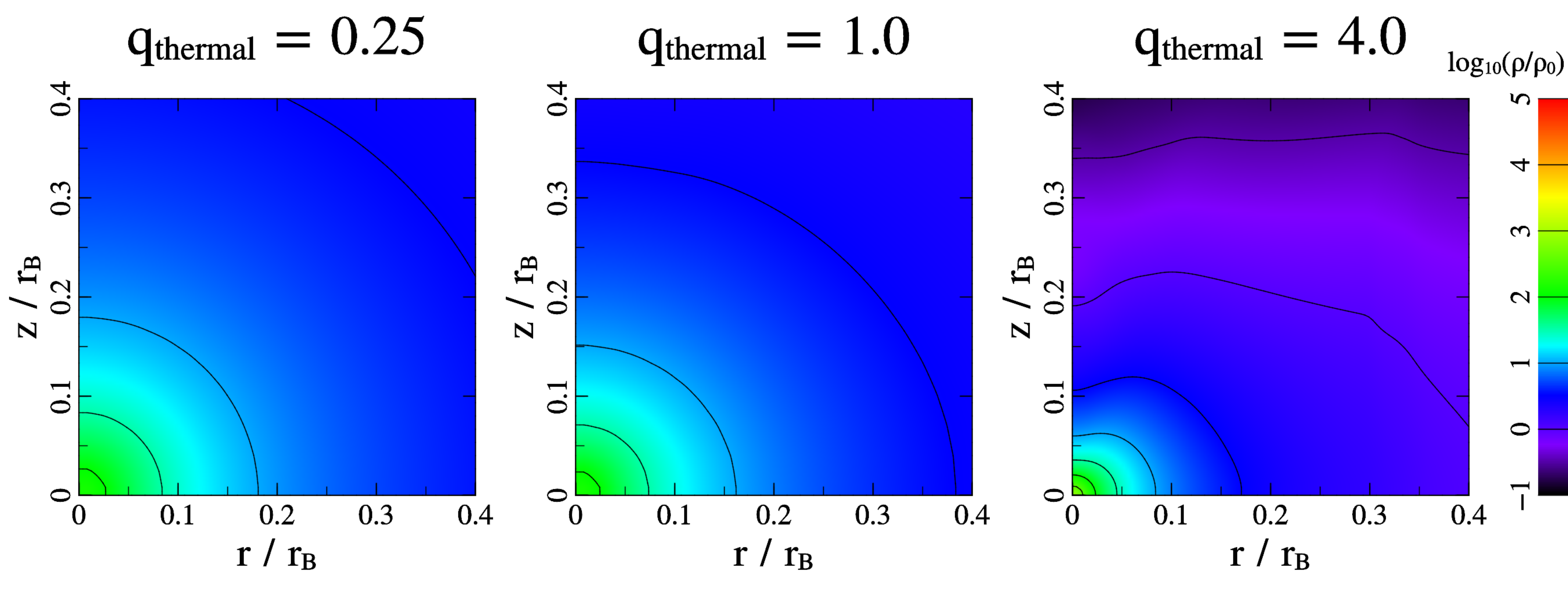}
\caption{Similar to \figref{fig:iso_den_2D}, but for adiabatic models \#8 (left), \#10 (middle), and \#12 (right). Compared to \figref{fig:iso_den_2D}, adiabatic CPDs are rounder and have lower densities. Subthermal adiabatic CPDs are nearly spherically symmetric. They become visibly flattened in the superthermal regime, but not as much so as the isothermal cases.}
\label{fig:ad_den_2D}
\end{figure*}

A simple and morphologically accurate description of our simulated CPDs is that isothermal CPDs are disks, while adiabatic CPDs are actually spheres (but we will continue to refer to them as circumplanetary ``disks,'' for convenience). 
Their typical morphologies are illustrated in \figref{fig:3D_surfaces}.

If one ignored rotation, one might expect CPDs to be in hydrostatic equilibrium. For isothermal gas, the spherically symmetric hydrostatic profile is:
\begin{equation}
\label{eq:iso_den}
\rho(r) = \rho_0 \exp{\left(\frac{\rB}{\sqrt{r^2 + \rs^2}}\right)}  \, ,
\end{equation}
while for an adiabatic gas with a constant entropy, the hydrostatic solution is:
\begin{equation}
\label{eq:poly_den}
\rho(r) = \rho_0 \left(\frac{\gamma-1}{\gamma}\frac{\rB}{\sqrt{r^2 + \rs^2}}-1\right)^{\frac{1}{\gamma-1}}  \, ,
\end{equation}
In both cases, we have included the effect of a smoothed gravitational potential using the smoothing length $\rs$, and $\rho_0$ is a normalization that is determined by the background density.
Our adiabatic simulations do not necessarily keep entropy constant, but because there are no shocks near the CPDs and the global entropy gradient is insignificant on the CPD scale, the entropy in our adiabatic CPDs is in fact roughly constant.
\figref{fig:ad_den_1D} compares the midplane density profile from 2 of our simulations to the polytropic profile described by \eqnref{eq:poly_den}, and they match very closely.

Isothermal CPDs, on the other hand, do not follow the exponential
hydrostatic profile. 
\figref{fig:iso_den_1D} shows that the midplane density profiles from our simulations are significantly more shallow. The \texttt{Athena++} profile shows a higher density than \texttt{PEnGUIn} inside $\sim0.05\rB$, which may be due to the use of a reflecting boundary condition at the inner boundary by \texttt{Athena++}; \texttt{PEnGUIn}, with a grid that is locally Cartesian close to the planet, has no boundary there. Nonetheless, the \texttt{Athena++} profile also deviates far from hydrostatic.
This difference between isothermal and adiabatic CPD structure is also demonstrated in \figref{fig:pressure}, where we see that in the adiabatic case, the pressure gradient is strong enough to balance the planet's gravity, but not when it is isothermal.

The fact that isothermal CPDs are not supported by gas pressure
raises the question of what is keeping it in steady state. One possibility is rotation. If the CPD rotates at Keplerian speed, that would provide the radial support it needs.
The other possibility is that it is in a steady-state flow. The gas can be constantly in motion, avoiding collapse by passing rapidly through the CPD. In fact, both ideas are correct --- part of the CPD is rotationally supported and the other part is constantly flowing in and out of the Bondi sphere. We will analyze gas flow in detail in the following sections.

The density structure has some dependence on $\qt$. \figref{fig:iso_den_2D} plots the r--z density structure of 3 isothermal models, and \figref{fig:ad_den_2D} plots the same for adiabatic models. Overall, the CPD structure scales well with $\rB$ for subthermal, $\qt\lesssim 1$ planets. The midplane radius at which the density becomes 10 times the background density, for example, is about $0.3$ to $0.4~\rB$ when isothermal. 
This value shrinks (in units of $\rB$) when we go to superthermal, $\qt\gtrsim 1$ planets; when $\qt=4$, it becomes $0.15~\rB$. 
This is perhaps not surprising. When $\qt>3$, the Hill radius is smaller than the Bondi radius, so one might expect the size to scale with $\rH$ instead.
Similarly for the adiabatic CPDs, their sizes, normalized by $\rB$, are about constant for subthermal planets, but shrink by a factor of 2 when going from $\qt=1$ to $4$. Moreover, when $\qt=4$, the adiabatic CPD becomes visibly flattened.

Below, we present our measurements of CPD sizes, rotation rates, and masses. We then describe a 3D view of the CPD flow structure. Finally, we discuss the effects of gap-opening on the CPD.

\subsection{Sizes}
\label{sec:size}

\begin{figure*}
\includegraphics[width=1.99\columnwidth]{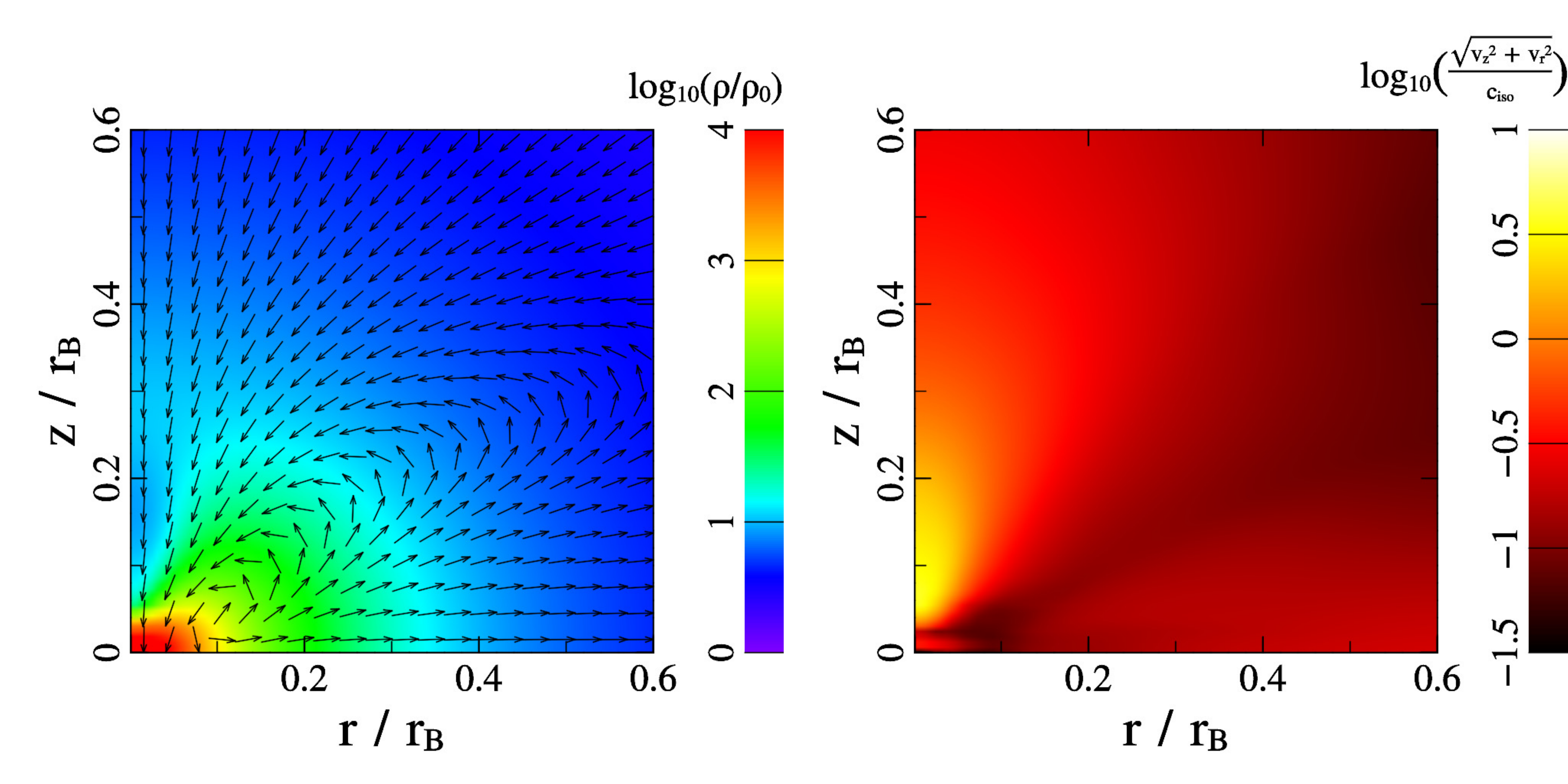}
\caption{Meridional flow pattern azimuthally averaged over $\phi$ for subthermal ($\qt\eqsim 1$) isothermal CPDs. The data is taken from model \#1. All subthermal and isothermal models share a similar pattern. On the left, arrow shows the direction of flow overplotted on top of the gas density in color. On the right, we show the magnitude of the meridional flow speed. The flow velocity directed downward toward the planet's pole reaches a maximum magnitude of about 4 to 5 times the sound speed. The midplane radial velocity (centered on the planet) changes sign at about $\sim0.1\rB$, indicating the CPD is bound within that distance.}
\label{fig:meri_iso}
\end{figure*}

\begin{figure*}
\includegraphics[width=1.99\columnwidth]{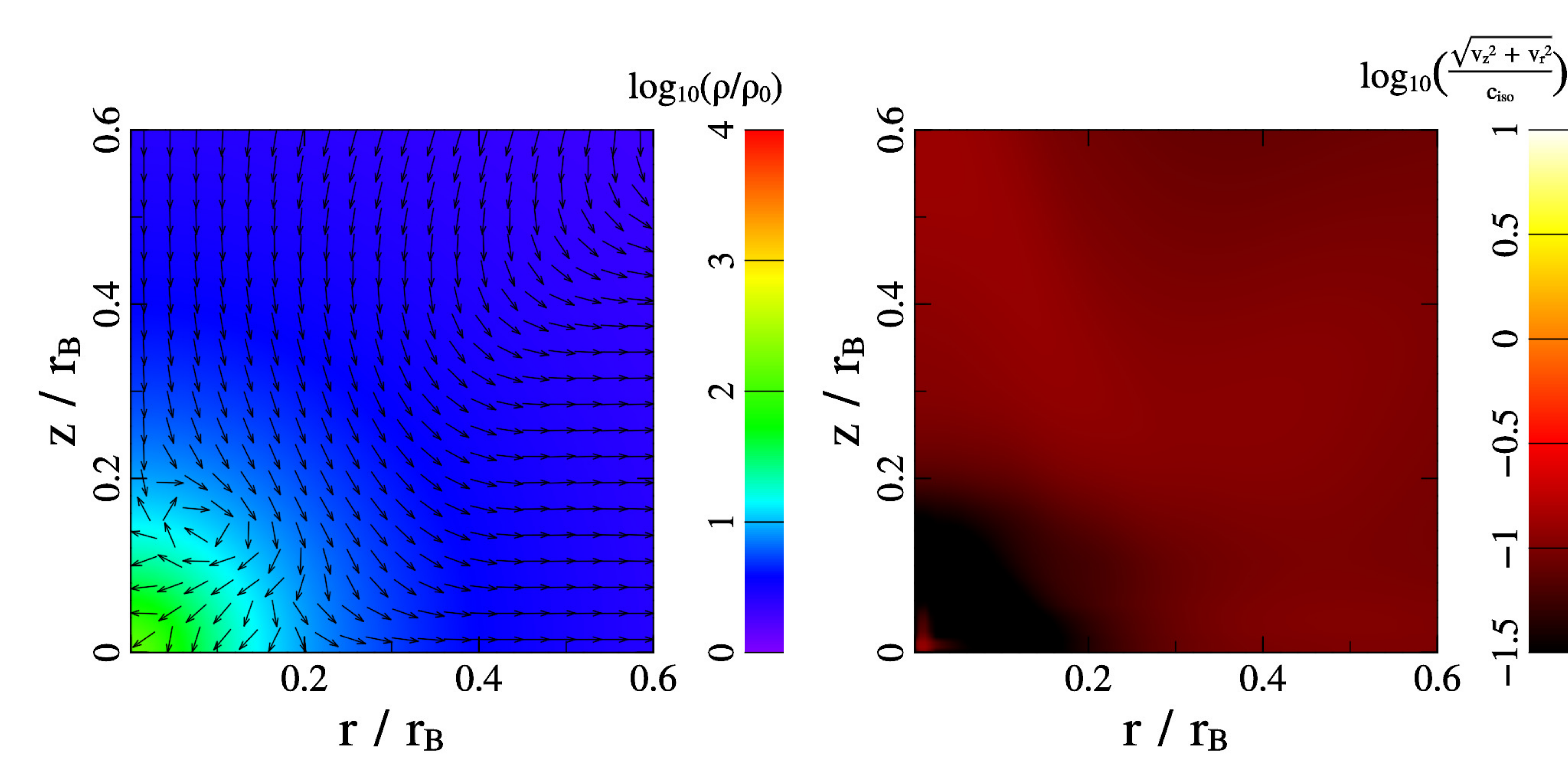}
\caption{Meridional flow pattern azimuthally averaged over $\phi$ for subthermal ($\qt\eqsim 1$) adiabatic CPDs. The data is taken from model \#7. All subthermal and adiabatic models share a similar pattern. Density and velocity scales are identical to the those used in \figref{fig:meri_iso} for ease of comparison. Unlike in the isothermal case, meridional flow speeds are subsonic everywhere, and the downward flow is deflected about $0.2\rB$ away from the planet. Within $0.2\rB$, flow speeds are substantially slower and the gas is likely bound to the planet.}
\label{fig:meri_ad}
\end{figure*}

\begin{figure}
\includegraphics[width=0.99\columnwidth]{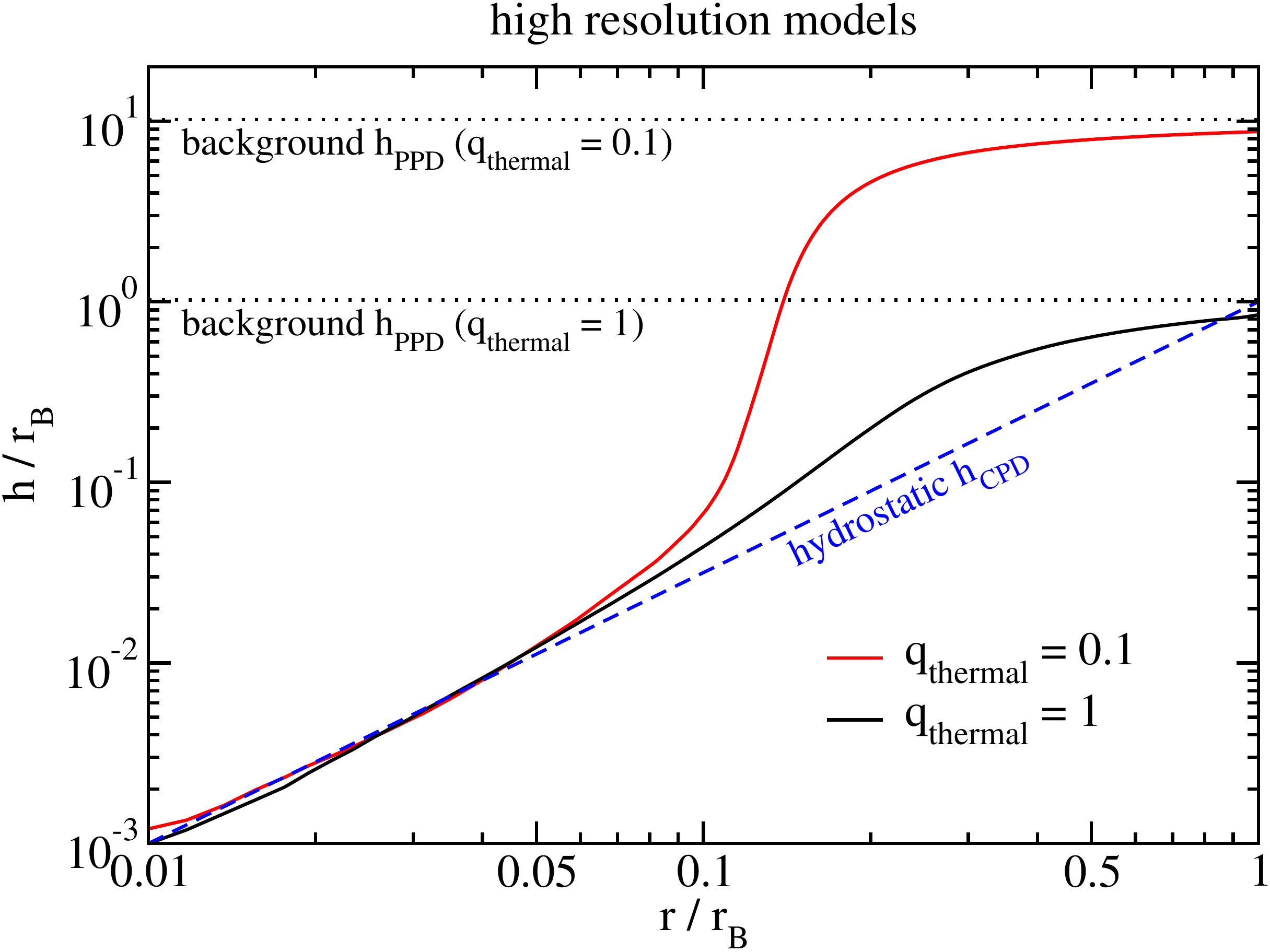}
\caption{Disk scale height vs.~cylindrical distance from the planet, showing results from the high resolution models, \#17 (black solid line) and \#18 (red solid line), 
with the scale height of the background disk $h_{\rm PPD}$
(black dotted) and the 
scale height of the CPD $h_{\rm CPD}$ 
(blue dashed) overlaid for comparison. Within $0.1\,\rB$, we find hydrostatic balance with the planet's gravity
($h/r_{\rm B} = h_{\rm CPD}/r_{\rm B} = (r/r_{\rm B})^{3/2}$). Outside $0.1\,\rB$, the disk puffs up and approaches the background scale height
($h/r_{\rm B} \approx h_{\rm PPD}/r_{\rm B} \propto h_{\rm PPD}/\qt$),
indicating that material is rapidly passing through the Bondi sphere and barely sensing the planet's gravity.}
\label{fig:hres_h}
\end{figure}

\begin{figure*}
\includegraphics[width=1.99\columnwidth]{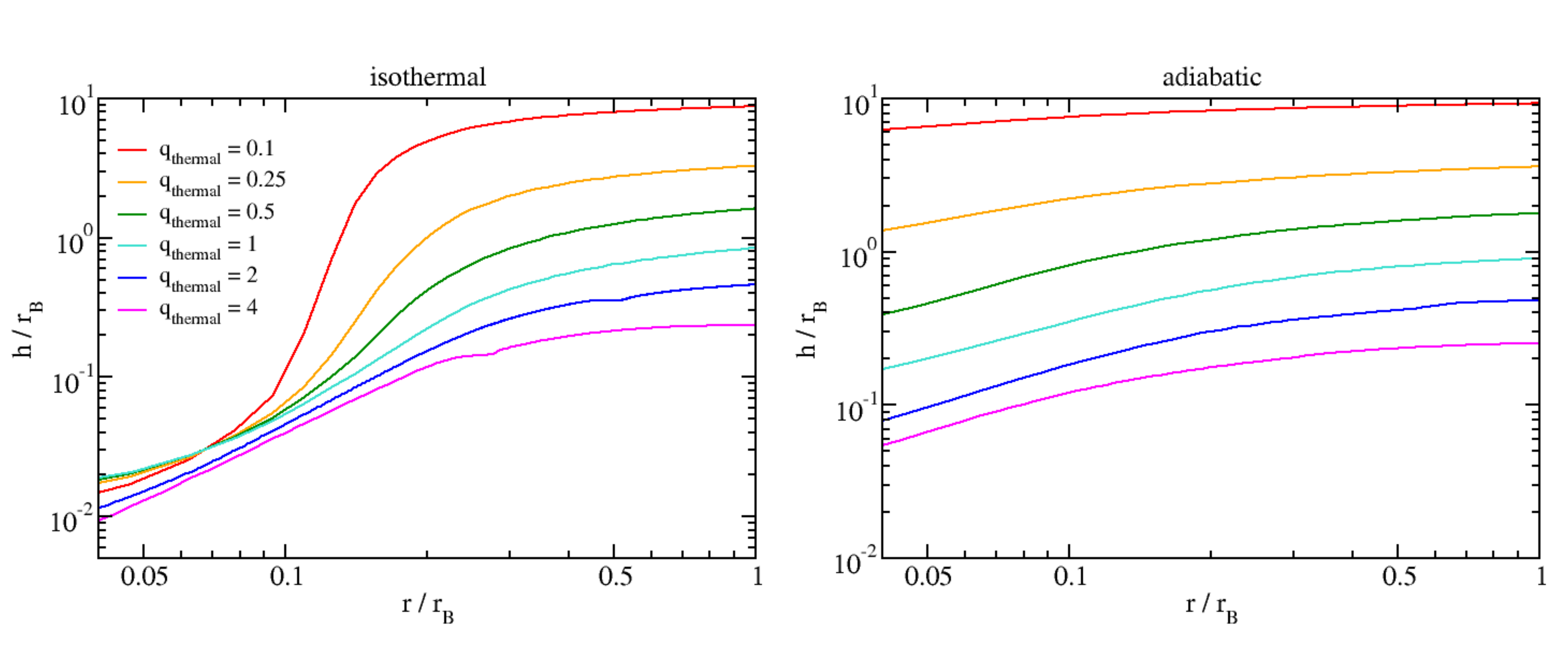}
\caption{Similar to \figref{fig:hres_h}, but from the fiducial resolution models. Left panel shows the isothermal models \#1--6, and right shows adiabatic models \#7--12. Isothermal results are in line with the high resolution ones in \figref{fig:hres_h}, although the lower resolution does lead to slightly larger scale heights. 
Unlike the isothermal profiles where different values of $\qt$ all converge to the same profile, a higher $\qt$ leads to a more midplane-concentrated vertical profile in the adiabatic models.}
\label{fig:h}
\end{figure*}

Determining which part of the gas is bound to the planet is nontrivial. For example, the specific energy of gas is not a good indicator, because it is not a conserved quantity. One way to infer boundedness is from kinematics. Gas that flows away from the planet is unbound; otherwise, it is bound and should be considered a part of the CPD.

Figures \ref{fig:meri_iso} and \ref{fig:meri_ad} show two 
examples of the azimuthally averaged meridional flow pattern, one each for the isothermal and adiabatic CPDs. Generally, gas flows vertically toward the planet along the poles, and away from the planet near the midplane, as has been shown in numerous previous studies \citep[e.g.,][]{Fung15,Ormel15,Bethune19,Kuwahara19}. This inflow is supersonic in the isothermal cases and can reach $4\sim5\,c_{\rm iso}$, but sub-sonic when adiabatic and generally does not exceed $0.1\,c_{\rm iso}$. In the midplane, the flow is directed away from the planet at larger distances, but toward the planet closer in. We can therefore use the location where the sign of the midplane radial velocity changes to characterize the sizes of our CPDs.

The inward flow occurs at around $\sim0.1\rB$ for isothermal CPDs, and $\sim0.2\rB$ for adiabatic ones. This measurement has some uncertainty because there are substantial temporal fluctuations in the velocity field close to the planet. In particular, the radial velocity can frequently change sign. Nonetheless, when averaged over time, we consistently find these specific locations to be where the radial velocity (centered on the planet) changes direction in all our models with $\qt\leq1$. For the largest $\qt$ tested, $\qt=4$, the isothermal CPD size is still about $0.1\,\rB$, but the adiabatic CPD size shrinks and becomes closer to $0.1\,\rB$. We expect the CPD size to eventually scale with $\rH$ instead of $\rB$ as $\qt$ increases, but our parameter space does not extend far enough to quantify that regime.

Another indicator of boundedness is the vertical density structure. If the gas is bound and has no significant vertical motion, it should settle into vertical hydrostatic equilibrium. The isothermal vertical hydrostatic density profile is:
\begin{equation}
\rho(r, z) = \rho_0(r) \exp{\left[\frac{\rB}{\sqrt{r^2 + z^2}}- \frac{\rB}{r}\right]} \exp{\left[-\frac{z^2}{2h_{\rm PPD}^2}\right]} \, .
\end{equation}
In the limit of $z\ll r$, it can be written as:
\begin{equation}
\rho(r, z) \approx \rho_0 \exp{\left[-\frac{z^2}{2}\left(\frac{1}{h_{\rm CPD}^2}+\frac{1}{h_{\rm PPD}^2}\right)\right]} \, ,
\end{equation}
where $h_{\rm CPD} = \sqrt{r^3 / \rB}$ is the expected scale height of the CPD if it is vertically settled.

The disk scale height as a function of distance from the planet is shown  in \figref{fig:hres_h} for our high-resolution isothermal models (\#17 and \#18); the same results but at fiducial resolution are shown in \figref{fig:h}, with isothermal runs (\#1--6) on the left and adiabatic runs (\#7--12) on the right. We measure the CPD scale height $h$ using the following definition:
\begin{equation}
0.68 = \frac{\int^{h}_{0} \rho~{\rm d}z}{\int^{z_{\rm max}}_{0}\rho~{\rm d}z} \, ,
\end{equation}
where $z_{\rm max}$ marks the top vertical boundary of the simulation. In other words, 68\% of the gas mass lies below $h$. This definition does not explicitly depend on the local sound speed and so can be used consistently in both the isothermal and adiabatic runs.

For isothermal CPDs, their vertical profiles follow the hydrostatic solution within $\sim0.1\rB$. This holds for all values of $\qt$ tested, as shown in the left panel of Figures \ref{fig:h} and \ref{fig:hres_h}. Beyond this distance, the disk expands vertically and 
follows the background scale height $h_{\rm PPD}$ instead. The gas outside $0.1\rB$ must therefore be unbound to the planet and passing through the Bondi sphere so rapidly that it barely reacts to the planet's gravity. This is consistent with the CPD sizes inferred from kinematics.

For the adiabatic runs, we measure small dips in $h$ directly above the planets, ranging from $\sim30\%$ for $\qt=0.1$ to a factor of $\sim4$ for $\qt=4$. These dips do not extend beyond $\sim0.2\rB$, which is again consistent with our interpretation that gas is unbound beyond that point.

When the local PPD aspect ratio is about 0.1, $\qt=0.1$ corresponds roughly to 0.1 Jupiter mass. Analyses of gaps in PPDs suggest that planets around this size may be common between 10 and 100 au \citep{Zhang18}. If these planets are present, our results indicate that their signature on the disk surface should be small if their CPDs are close to adiabatic. This may explain why they are not observed directly, despite their prominent gaps.

\subsection{Rotation}
\label{sec:rotation}

In the classical 2D picture, 
the background Keplerian shear provides the source of angular momentum for CPDs. Gas is accreted by the planet through the L1 and L2 Lagrange points; material enters the Hill sphere with an angular momentum of roughly $\rH^2 \Omega_{\rm p}$. Setting this equal to the Keplerian angular momentum around the planet $\sqrt{GM_{\rm p}r}$, one finds $r=\rH/3$, implying that the CPD is rotationally supported within $\sim\rH/3$ \citep{Quillen98}. 2D calculations focusing on the effects of tidal truncation produced a similar disk size \citep{Martin11}.

This picture is modified significantly in 3D. Previous studies have found that, in 3D, planets accrete gas from the vertical direction instead. That gas originates directly above the planet and could be co-orbiting with it. A small orbital velocity difference between the gas and the planet would mean the gas has a lower angular momentum than in 2D CPDs.

How low might this angular momentum be? The lower it is, the smaller the rotationally supported region is, and the higher the required resolution becomes. This presents a challenge to our ability to resolve the CPD. In \secref{sec:resolution}, we have demonstrated that our fiducial resolution is converged for the rotation speed on scale $\sim\rB$, but it may not be sufficient if the rotationally supported region turns out to be much smaller than $\rB$. We shall bear this in mind as we proceed.

The Keplerian rotation around a planet can be expressed in terms of $c_{\rm iso}$ and $r_{\rm B}$:
\begin{equation}
\label{eq:kep_vel}
\frac{v_{\rm K}}{c_{\rm iso}} = \left(\frac{r}{\rB}\right)^{-1/2} \, ,
\end{equation}
as long as we use $c_{\rm iso}$ when defining $\rB$ (\eqnref{eq:rB}), and the specific angular momentum profile is similarly:
\begin{equation}
\label{eq:kep_ang}
\frac{l_{\rm K}}{\rB c_{\rm iso}} = \left(\frac{r}{\rB}\right)^{1/2} \, .
\end{equation}
We will normalize the speeds and distances of our results by $c_{\rm iso}$ and $r_{\rm B}$ to directly compare simulations with different planet and disk parameters. \figref{fig:rot} plots the midplane rotation profiles from the isothermal simulations (models \#1--6) in the left panel, and the adiabatic results (models \#7--12) are on the right. These profiles are azimuthally averaged in the 
planet-centered
frame, but we note that within $\rB$ there generally is little azimuthal variation.

\subsubsection{Isothermal Disks}
\label{sec:rotation_iso}

\begin{figure*}
\includegraphics[width=1.99\columnwidth]{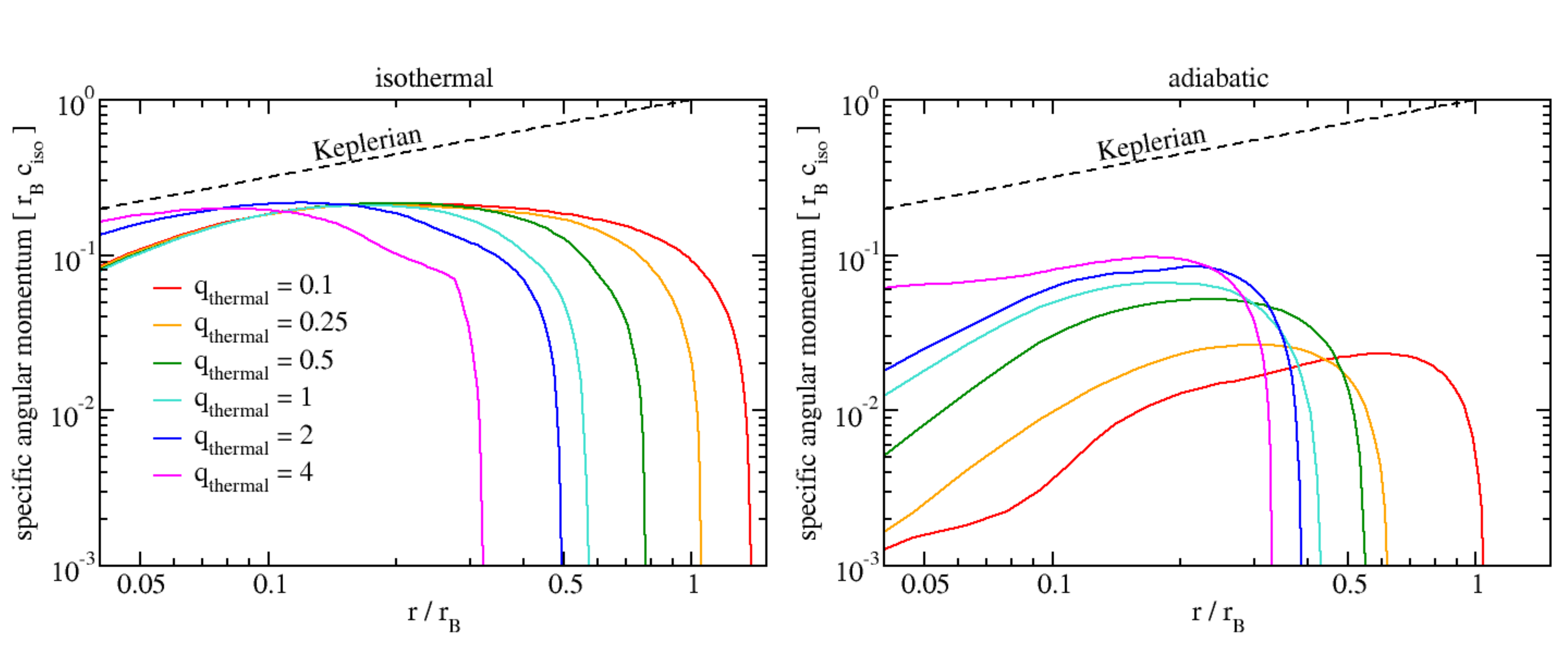}
\caption{Midplane rotation curves azimuthally averaged over $\phi$ for different values of $\qt$. Rotation speeds are normalized by the isothermal sound speed, and distances from the planets by $\rB$. Left panel shows the isothermal simulations (models \#1--6), and right shows adiabatic ones (models \#7--12), all at our fiducial resolution. The Keplerian profile is shown as the black dashed line. To make a fair comparison, the speeds in the right panel are normalized by the same isothermal sound speed (not the adiabatic sound speed) as those on the left.}
\label{fig:rot}
\end{figure*}

\begin{figure*}
\includegraphics[width=1.99\columnwidth]{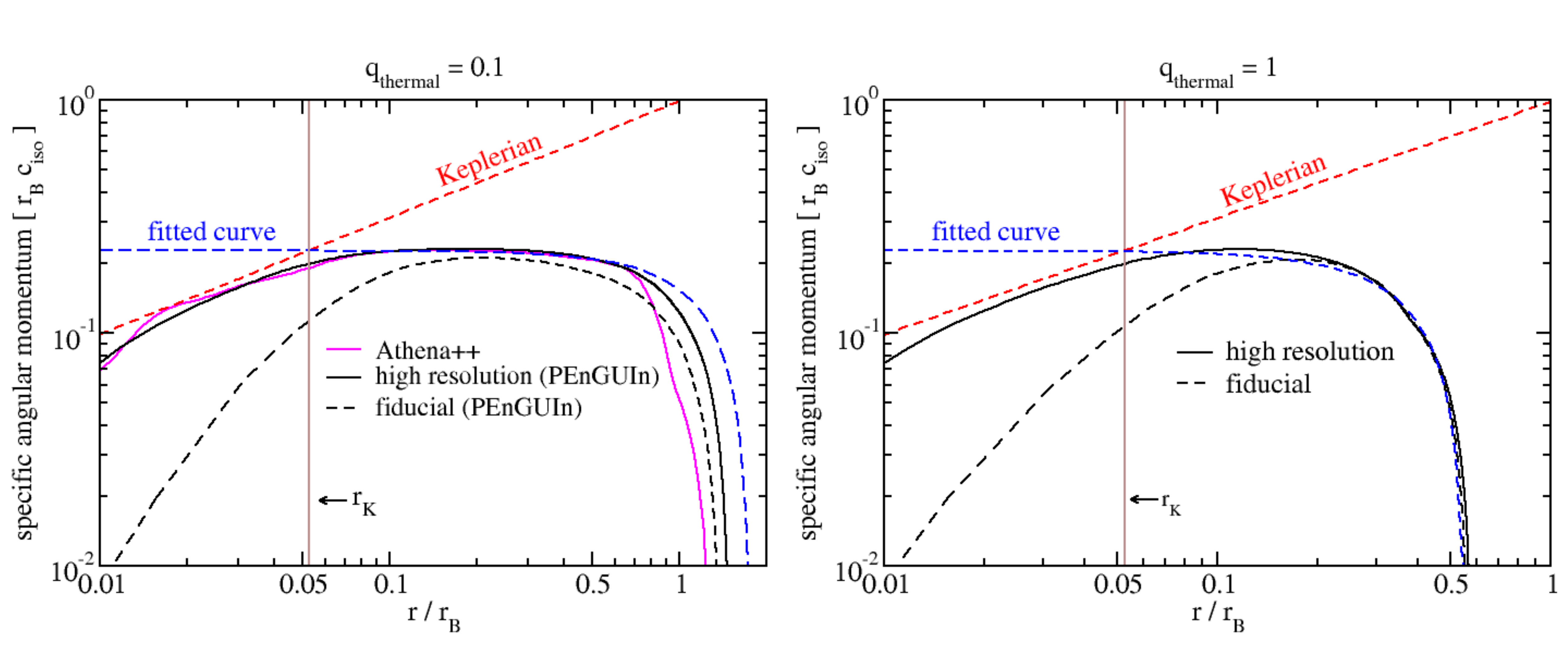}
\caption{Similar to \figref{fig:rot}, the black solid lines are the specific angular momentum profiles from \texttt{PEnGUIn} simulations with 8 times higher resolution near the planet (models \#17 and \#18), and the magenta solid line on the left panel is an \texttt{Athena++} simulation. We also show the fiducial simulations in black dashed lines for comparison. Red dashed lines correspond to Keplerian rotation, and blue dashed lines are the fitted rotation curves described by \eqnref{eq:iso_fit}. Our fits closely match our simulations from 0.05 to 1 $\rB$. Inside $r_{\rm K}\sim0.05\,\rB$ it transitions to Keplerian rotation, as demonstrated by our high resolution simulations.}
\label{fig:r_K}
\end{figure*}

\begin{figure}
\includegraphics[width=0.99\columnwidth]{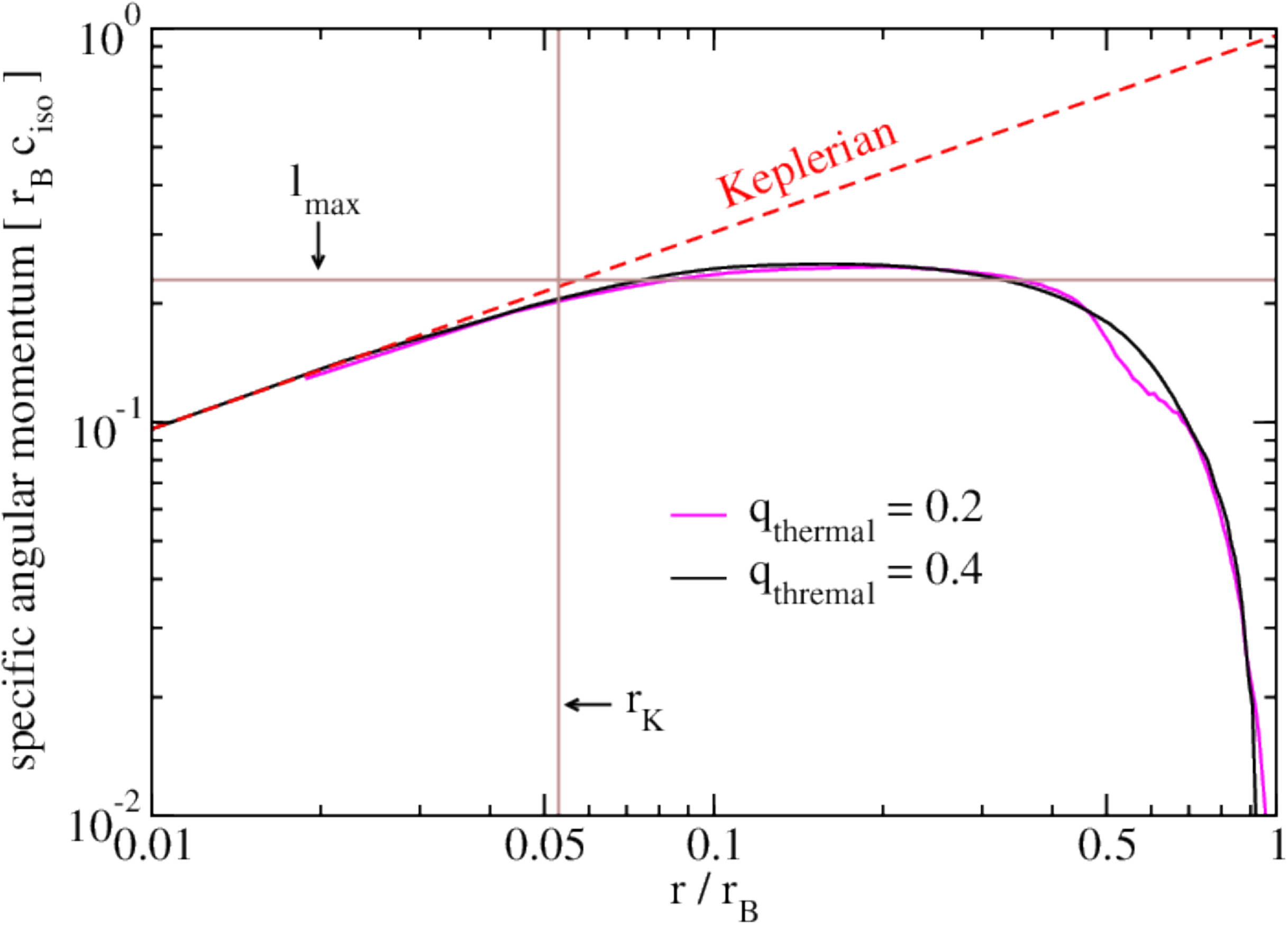}
\caption{Same as \figref{fig:r_K}, but with simulation data taken from \citet{Wang14}, courtesy of Chun-Fan Liu and Hsien Shang. Their rotation curves agree well with our fits, and similarly show Keplerian disks of sizes about $0.05\,\rB$.}
\label{fig:W14}
\end{figure}

\begin{figure*}
\includegraphics[width=1.99\columnwidth]{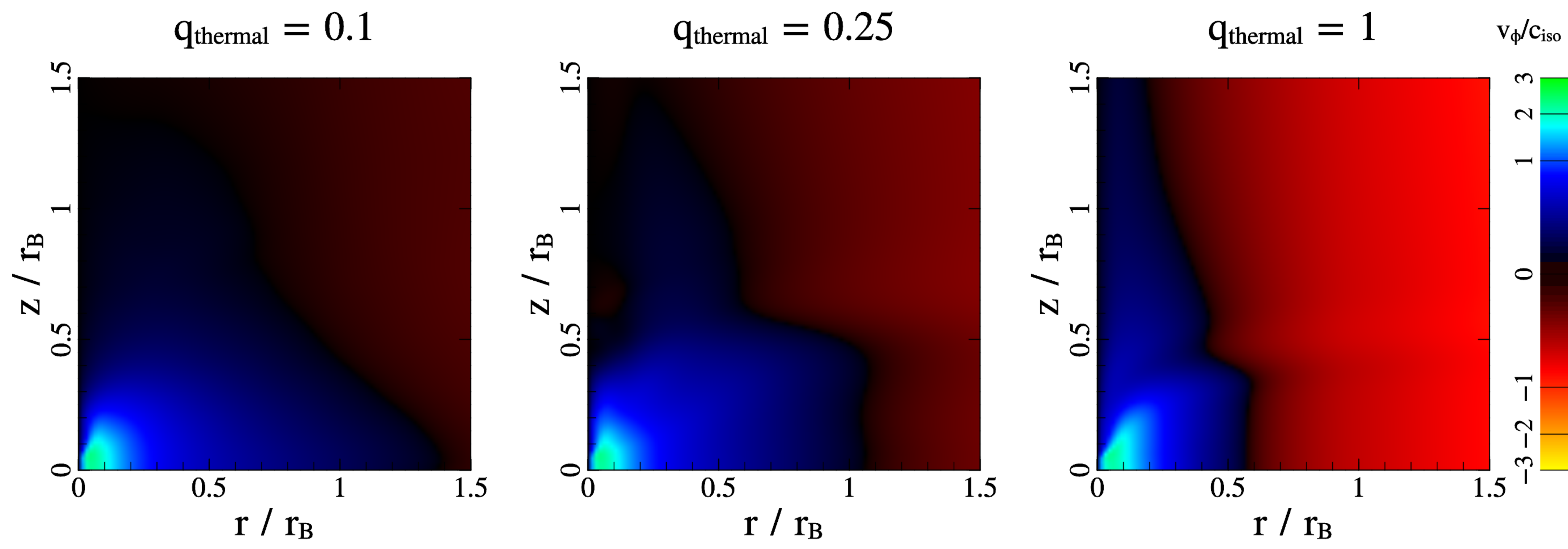}
\caption{Rotation speeds from the isothermal models \#1, 2, and 4. The data is azimuthally averaged over $\phi$ and shown as a function of radius and vertical height. Blue (positive) indicates prograde rotation; red (negative) is retrograde and corresponds to the background Keplerian shear. The rotation structure is more columnar for higher planet masses.}
\label{fig:rot_rz}
\end{figure*}

Strikingly, within a distance of $\sim0.2,\rB$, all angular momentum profiles from various $\qt$ converge to a single value in units of $\rB c_{\rm iso}$. For $0.1\leq\qt\leq1$, the profiles nearly lie on top of each other. The $\qt=$2 and 4 simulations have a shorter normalized smoothing length $\rs/\rB$ (see \tabref{tab:para}), and so they reach higher speeds at very short distances . 

The fact that we can achieve higher rotation speeds by decreasing $\rs/\rB$ suggests that $\rs$ still has too large of an effect on the rotation at our fiducial resolution. We therefore increase the resolution inside $\sim 0.2 \rB$ by a factor of 8, correspondingly reducing $\rs$ by a factor 8, to produce models \#17 and \#18. Figure \ref{fig:r_K} plots the angular momentum profiles from those models. They confirm that rotationally supported, Keplerian disks indeed exist around these planets.

The angular momentum profiles around subthermal planets can be approximated as a superposition of a constant value and the background shear that one would obtain in the absence of the planet. A formal fit gives:
\begin{equation}
\label{eq:iso_fit}
\frac{l}{\rB c_{\rm iso}} = \frac{l_{\rm max}}{\rB c_{\rm iso}}~ - ~\frac{3\qt}{4}\left(\frac{r}{\rB}\right)^2\, ,
\end{equation}
where $l_{\rm max} = 0.23\,\rB c_{\rm iso}$ and the second term on the right corresponds to the background shear. We overplot this profile in Figure \ref{fig:r_K} to show that it compares well with our empirical profiles.

The value $0.23\,\rB c_{\rm iso}$ corresponds to the Keplerian angular momentum at about $0.05\,\rB$. This is the size of the Keplerian disk, which we denote as $r_{\rm K}$. 
The scaling $\rB c_{\rm iso}$ can alternatively be expressed as $x_{\rm s}^2 \Omega_{\rm p}$, where $x_{\rm s}$ takes the subthermal branch in \eqnref{eq:xs}. Therefore, $l_{\rm max}$ may be related to the incoming momentum in the horseshoe orbits, which is consistent with that idea that the CPD is fed by the horseshoe flow \citep{Fung15}.
Even though our parameter space only covers down to $\qt=0.1$, given the lack of dependence on $\qt$ in the rotation profile, we expect our results to apply to all subthermal planets. This implies that under isothermal conditions, even smaller planets (such as the Earth) should have rotationally supported CPDs inside $r_{\rm K} \approx 0.05 \rB$.

For superthermal planets, the maximum angular momentum in their CPDs is also about $0.23\rB c_{\rm iso}$, but the size of the region with this specific angular momentum rapidly shrinks with increasing $\qt$, to the point that the overall profile significantly deviates from \eqnref{eq:iso_fit}. The left panel of Figure \ref{fig:rot} shows that when $\qt=4$, $l_{\rm max}$ is reached at just about $0.15 \rB$. 

In this regime, it is likely that we are beginning to see the transition of the CPD from being limited by the Bondi radius to the Hill radius. Clearly, one should not expect $l_{\rm max}$ to scale with $\rB$ indefinitely, or else the size of the $r_{\rm K}$ disk will eventually exceed $\rH$.
One commonly suggested scaling for superthermal planets is $r_{\rm K}\sim \rH/3$. If we take that scaling, then the transition would occur near $\rH/3 = \rB/20$, corresponding to $\qt\sim10$.

To address the possibility of code bias, we also compare our inferred values of $l_{\rm max}$ and $r_{\rm K}$ to the simulations by \citet{Wang14}, who used the \texttt{Antares} code and its static mesh refinement to attain resolutions comparable to our models \#17 and \#18. \figref{fig:W14} plots the angular momentum profiles from two of their models with subthermal planet masses of $\qt=0.2$ and $0.4$.
Their results agree with ours; the maximum angular momentum from their profiles differ from ours by about 10\%, and their Keplerian disk sizes are also similar. We emphasize that the three codes we have used for comparison, \texttt{PEnGUIn}, \texttt{Athena++}, and \texttt{Antares}, all used different setups: \texttt{PEnGUIn} uses a nonuniform spherical grid centered on the star, \texttt{Athena++} uses a logarithmic spherical grid centered on the planet, and \texttt{Antares} uses a cylindrical grid with mesh refinement centered on the star. Additionally, a similar isothermal simulation carried out by \citet{Tanigawa12} also found a maximum specific angular momentum of about $0.2\,\rB c_{\rm iso}$ (expressed in their units as $0.7\,\rH^2 \Omega_{\rm p}$) for a $\qt=3$ planet. The agreement between all these results lends confidence to our findings.

Similar experiments have been performed by \citet{Ormel15}. They simulated planets with $\qt=0.01$ and report that there is negligible rotation in the CPD. Their simulation domain extends as close to the planet as about $0.055\,\rB$. Since this is similar to $r_{\rm K}$, the region where we expect to see Keplerian rotation is cut out from their domain. Their Figure 4, top panel, suggests that if they had set their inner boundary smaller, they would have seen faster rotation.

We also inspect how the rotation rate changes vertically. \figref{fig:rot_rz} plots the azimuthally averaged r--z rotation profiles from models \#1, 2, and 4. The speed is slower at higher altitudes, but the overall prograde rotation pattern does extend vertically to about $1\,\rB$. At higher planet masses, rotation in the CPD appears to become more columnar.

\subsubsection{Adiabatic Envelopes}
\label{sec:rotation_ad}

The right panel of \figref{fig:rot} tells a much different story for adiabatic CPDs. Unlike the isothermal cases, we do not find speeds close to the Keplerian value in any of our simulations. Even at $\qt=4$, the rotation speed reaches only about one-third of the Keplerian speed.

Also unlike the isothermal CPDs, where rotation profiles follow a similar pattern regardless of $\qt$, adiabatic CPDs increase in rotation speed as $\qt$ increases. This trend can be understood as an effect of the Coriolis force. Because the adiabatic gas around the planet is, as we have seen in \figref{fig:ad_den_1D}, roughly in hydrostatic equilibrium, vertical gas flow toward the planet must be deflected and turned to planar motion as it flows over this ball of hydrostatic gas. In the planet's frame, the deflected gas must then be torqued by the Coriolis force into prograde rotation. Assuming the bound atmosphere has a size of $C\rB$, where $C$ is a scaling coefficient, the rotation speed by the time the gas reaches the equator plane is approximately:
\begin{equation}
\label{eq:ad_vphi}
v_{\phi} \approx v_{\rm gas} \Omega_{\rm p} \frac{C\rB}{v_{\rm gas}} \, ,
\end{equation}
where $v_{\rm gas}$ is the meridional speed of the gas as it gets deflected around and flows over the planet's atmosphere, $v_{\rm gas} \Omega_{\rm p}$ is the Coriolis force (dropping the factor of 2), and $C\rB/v_{\rm gas}$ is the timescale of the deflection (also dropping order-unity prefactors). We can combine this expression with \eqnref{eq:kep_vel} to scale it with the Keplerian speed. At the atmosphere's boundary, $r=C\rB$, the rotation speed as a fraction of the Keplerian speed is then:
\begin{equation}
\label{eq:ad_rot_sup}
\frac{v_{\phi}}{v_{\rm K}} \approx C^{3/2} \frac{\rB \Omega_{\rm p}}{c_{\rm iso}} = C^{3/2} \qt \, .
\end{equation}
This fraction scales linearly with $\qt$, in rough agreement with our results in the right panel of \figref{fig:rot}. 
We have seen that adiabatic CPDs are approximately bound within $0.2\,\rB$.
Plugging this into \eqnref{eq:ad_rot_sup}, we get $v_{\phi}/v_{\rm K}\sim 0.1 \qt$, which is in good quantitative agreement with our results. Furthermore, this suggests that adiabatic CPDs can potentially become rotationally supported if $\qt\gtrsim10$.

Our adiabatic results can be roughly compared to simulations where effects of radiative transfer are included, such as those by \citet{Szulagyi16}, \citet{Szulagyi17}, \citet{Cimerman17}, and \citet{Lambrechts17}. Before the gas can cool significantly, it is roughly adiabatic and is comparable to our simulations. Our results are similar to those by \citet{Cimerman17} and \citet{Lambrechts17}, who also found little to no rotation in their circumplanetary gas. We agree qualitatively with \citet{Szulagyi16} and \citet{Szulagyi17}, in that they measured slower rotation when the gas cools more slowly, but we note that their simulations use large values of $\qt$ ranging from 8 to 80, significantly different from our parameter space.

\subsection{Masses}
\label{sec:mass}
\begin{figure*}
\includegraphics[width=1.99\columnwidth]{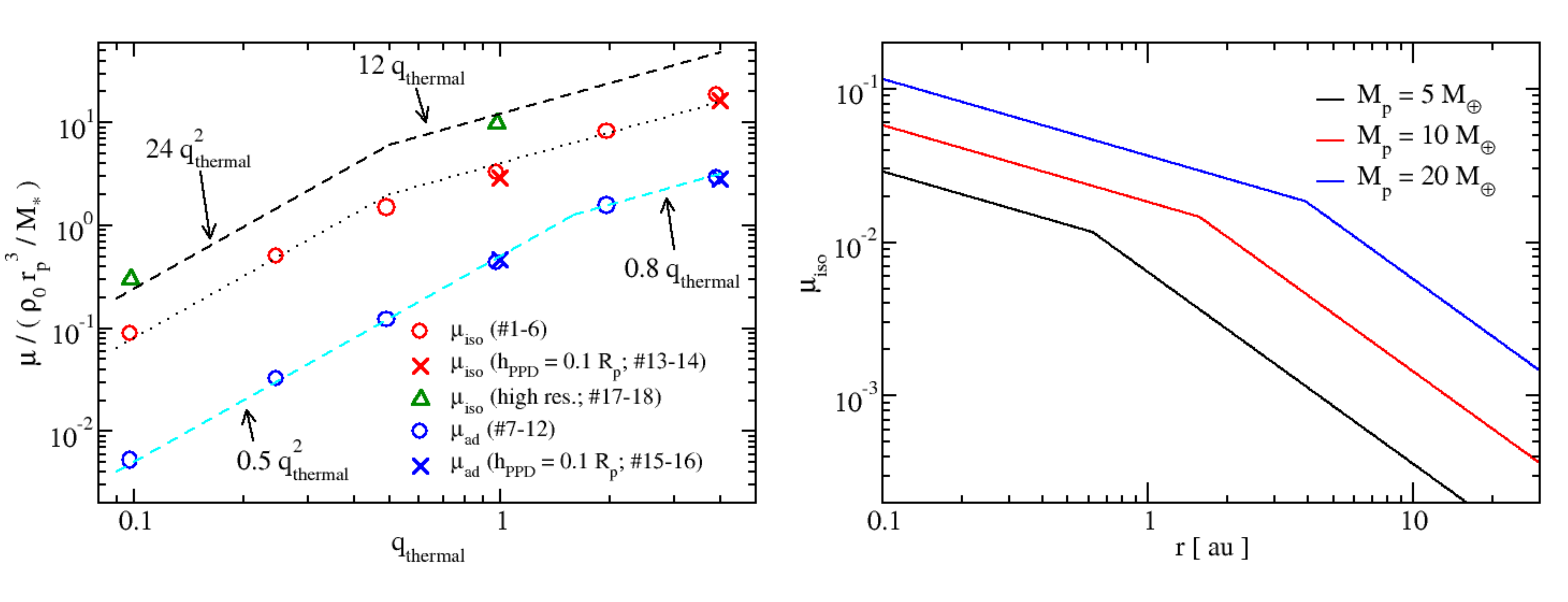}
\caption{The gas-to-core mass ratios, $\mu$, as functions of $\qt$ (left) and orbital radius in a minimum-mass solar nebula (right). On the left, we also plot an approximate fit to our data points at fiducial resolution for isothermal CPDs (black dotted lines) and adiabatic CPDs (cyan dashed lines). Isothermal results are more sensitive to resolution, so we also re-normalize our isothermal fit to the high resolution points (black dashed lines). For subthermal planets, $\mu$ scales with $\qt^2$, implying the CPD mass $M_{\rm CPD}$ scales with $\rB^3$. For superthermal planets, it instead scales linearly with $\qt$, consistent with $M_{\rm CPD}\sim\rB^2 h_{\rm PPD}$. The dividing point is around $\qt=0.5$ for isothermal runs and $\qt=1.6$ for adiabatic ones. Adiabatic CPDs are less massive then isothermal ones, by a factor of $\sim 48$ for subthermal planets and $15$ for superthermal ones. On the right, we use the high resolution fit to estimate $\mu_{\rm iso}$ in an MMSN for 3 different core masses within the super-Earth range. A super-Earth between 0.1 and 1 au has a fully cooled planetary atmosphere weighing a few percent of the core's mass.}
\label{fig:mu}
\end{figure*}

\begin{figure*}
\includegraphics[width=1.99\columnwidth]{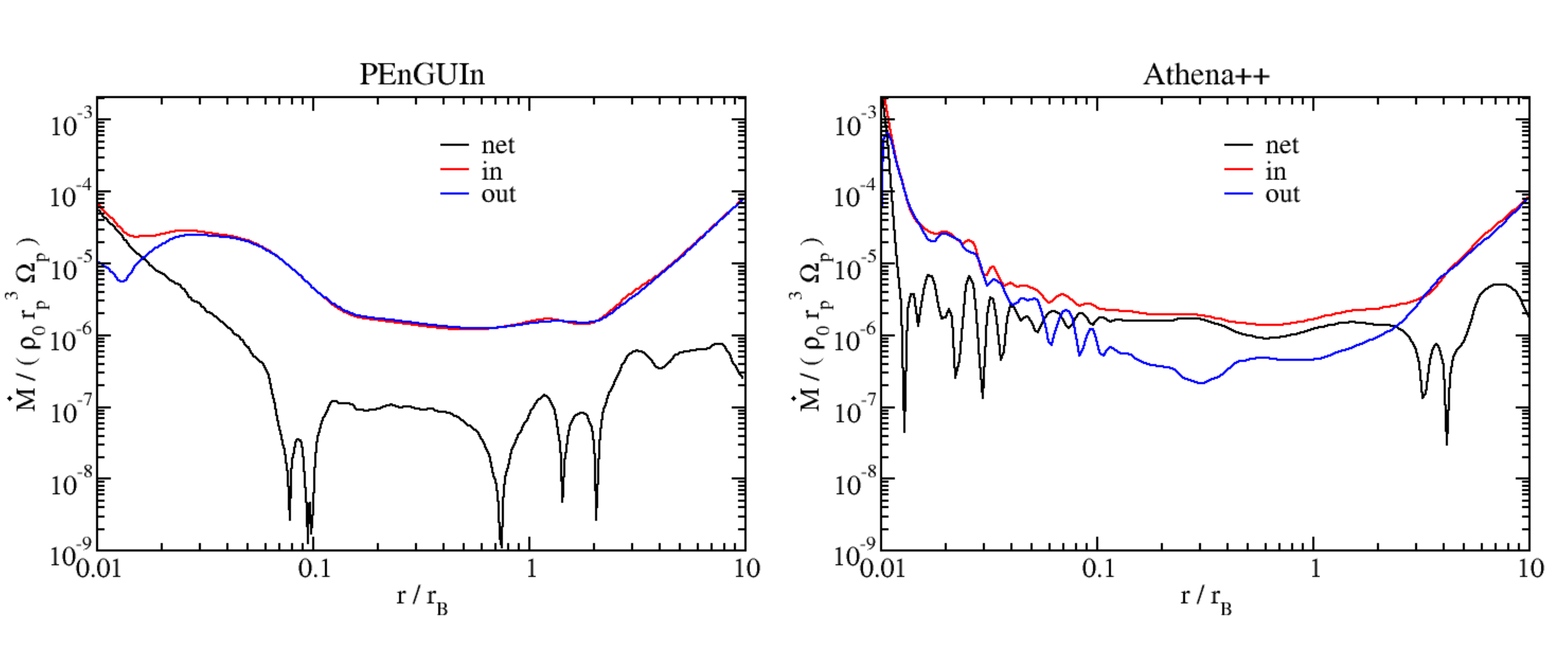}
\caption{The magnitudes of mass fluxes, $|\dot{M}|$, as functions of distance from the planet. Red curves are influxes toward the planet, blue are outfluxes away from the planet, and black are the net fluxes. Both panels plot results from model \#17, with the left showing results from \texttt{PEnGUIn} and right from \texttt{Athena++}. Both are snapshots at the end of the simulations without any time averaging. In \texttt{PEnGUIn}, the in and out fluxes balance to within a few percent inside most of the domain---except within $\sim 0.03~\rB$, where we find some accretion. In \texttt{Athena++}, there is less outflux, which leads to overall accretion across the entire Bondi radius. Comparing to \figref{fig:mu}, we find that $M_{\rm CPD}/\dot{M}\sim 10^{3}-10^{4}~\Omega_{\rm p}^{-1}$, which is much longer than our simulation time. This suggests the density structures have largely settled to a steady state.}
\label{fig:mflux}
\end{figure*}

Isothermal models are commonly interpreted to represent the final state of the protoplanet's atmosphere, after it has fully cooled to the background nebular temperature
(e.g., \citealt{Lee15}, their figure 4 and related discussion). In 1D, spherically symmetric models, such a final state would be described by the hydrostatic profile of \eqnref{eq:iso_den}. In 3D hydrodynamical simulations, we have seen that the density profile differs (\figref{fig:iso_den_1D}), having much lower densities. It seems clear, then, that 1D models overestimate the gas mass in isothermal planetary atmospheres, and to this extent may overpredict the likelihood of giant planet formation. Here, we measure the total gas mass, $M_{\rm CPD}$, in our CPDs and
compute the gas-to-core mass ratios, $\mu\equiv M_{\rm CPD}/\Mp$.

We compute $M_{\rm CPD}$ by summing the gas mass within a sphere of $0.1~\rB$ for the isothermal cases or $0.2~\rB$ for the adiabatic cases,
following the CPD sizes measured in \secref{sec:size}.
Since we only simulate half the disk and assume midplane symmetry, we multiply the sum total by 2 to get the full $M_{\rm CPD}$.
A note about units: the mass so computed is scaled to the ambient nebular gas density $\rho_0$, and 
has units of $\Mstar$. Because the code takes $\rho_0=1$ in units of $\Mstar/\rp^3$
(the exact value is immaterial because gas self-gravity is neglected), to scale to any other nebular
density $\rho_0$ we multiply by $\rho_0 / (\Mstar/\rp^3)$. Then to convert into physical units,
we multiply by $\Mstar$. In sum, to convert $M_{\rm CPD}\, ({\rm code\, units})$ into $M_{\rm CPD}$ in
physical units, we compute $M_{\rm CPD} = M_{\rm CPD} \, ({\rm code \, units}) \times \rho_0 \rp^3 / \Mstar \times \Mstar = M_{\rm CPD} \, ({\rm code \, units}) \times \rho_0 \rp^3$.

We find the gas-to-core mass ratios $\mu$ in both
our isothermal and adiabatic simulations
to scale the same way with $\qt$.
For subthermal planets, not surprisingly, we find $M_{\rm CPD}$ to scale with the volume of the Bondi sphere, $\rB^3$, which implies $\mu  \propto \rB^3/\Mp \propto \Mp^2 H_{\rm p}^{-6} \propto \qt^2$. This scaling breaks down when we reach superthermal masses, where $\mu$ starts to scale linearly with $\qt$ instead; this can be understood as $M_{\rm CPD}\propto\rB^2 h_{\rm PPD}$, where $h_{\rm PPD} = \rp H_{\rm p}$ is the protoplanetary (circumstellar) disk scale height. The left panel of \figref{fig:mu} shows these scalings match well with our measurements.

For isothermal CPDs, $\mu = \mu_{\rm iso}$ is about 3 times higher in our high resolution models
compared to our fiducial ones. 
Because we trust the high resolution results more, but also because the lower resolution fiducial simulations better sample parameter space, we use the fiducial models to guide our scaling, and the high resolution models to normalize these scalings. Our final, empirical measurement of $\mu_{\rm iso}$ is:
\begin{equation}
    \mu_{\rm iso} = \begin{cases}
        24~\qt^2 \times \rho_0 \rp^3 / \Mstar  \, , & \qt\leq 0.5 \, ; \\
        12~\qt \times  \rho_0 \rp^3 / \Mstar  \, , & \qt>0.5 \, ,
    \end{cases}
    \label{eq:mu_iso}
\end{equation}
where the multiplicative factor allows us to scale to any desired background nebular density $\rho_0$
(see above note about units). 
The \texttt{Athena++} simulation for model \#17 produces a higher $\mu_{\rm iso}$ than \texttt{PEnGUIn} by one order of magnitude. This discrepancy can also be seen in \figref{fig:iso_den_1D}. As mentioned in the beginning of \secref{sec:result}, this is likely due to mass accumulating in front of the reflecting boundary used by \texttt{Athena++}. The effects of different boundary conditions at the planetary core need to be investigated further in the future. \texttt{PEnGUIn} has no boundary at the planet's location, so its results are easier to interpret.

Figure \ref{fig:mflux} further illustrates this difference between \texttt{PEnGUIn} and \texttt{Athena++} and investigates the steadiness of our CPDs. It plots the mass flux, $\dot{M}$, across the sphere at a given distance $r$ centered on the planet. \texttt{PEnGUIn} shows a close balance between the in (toward the planet) and out (away from the planet) fluxes, while the influx dominates in \texttt{Athena++}, resulting in a higher accretion rate. Despite the difference, the net $\dot{M}$ values are small in both cases. We get roughly $M_{\rm CPD}/\dot{M}\sim 10^3-10^4~\Omega_{\rm p}^{-1}$, which we consider nearly steady. A similar level of steadiness is found in all of our models.

We emphasize that our measurements of $\mu_{\rm iso}$ are 
many orders of magnitude below what they would be if the gas were to follow the 1D hydrostatic profile described by \eqnref{eq:iso_den}.
This holds true for both \texttt{PEnGUIn} and \texttt{Athena++} results, and 
represents one of the most
important differences between 1D models and 3D hydrodynamics simulations.

For adiabatic CPDs, resolution is
less of a concern. Since at fiducial resolution, adiabatic density profiles closely follow the 1D hydrostatic solution, we believe the numerical solution to have converged. A fit to our measured values of $\mu_{\rm ad}$ is given by
\begin{equation}
    \mu_{\rm ad} = \begin{cases}
        0.5~\qt^2 \times \rho_0 \rp^3 / \Mstar \, , & \qt\leq 1.6 \, ; \\
        0.8~\qt \times \rho_0 \rp^3 / \Mstar \, , & \qt>1.6 \, .
    \end{cases}
    \label{eq:mu_ad}
\end{equation}
This result is more sensitive to the choice of CPD size, because adiabatic CPDs are much less centrally concentrated than isothermal ones. If we had used $1~\rB$, for example, instead of $0.2~\rB$, then $\mu_{\rm ad}$ would increase by about an order of magnitude. 
This is not concerning because we have seen evidence 
that the gas beyond $\sim0.2~\rB$ is unbound (\secref{sec:size}).

We can estimate numerical values for $\mu_{\rm iso}$ for real-world applications. For this we use the same disk density profile as in \eqnref{eq:init_den}, and set $\rho_0 = \Sigma_{\rm MMSN}/\sqrt{2\pi h_{\rm PPD}^2}$, where $\Sigma_{\rm MMSN} = 1700 \, {\rm g~cm^{-2}}$ is the surface density of the minimum-mass solar nebula (MMSN) at 1 au. For the temperature profile, we choose one such that
$h_{\rm PPD}/R = 0.035~(R/1~{\rm au})^{1/4}$. For $\Mstar$, we use one solar mass.
Plugging these values into \eqnref{eq:mu_iso}, we show in the right panel of \figref{fig:mu} the gas-to-core mass ratios of planets with core masses ranging from 5 to 20 $\Me$. Within a few au, the cores are superthermal and $\mu_{\rm iso}$ is on the order of a few percent. Outside a few au, the cores are subthermal and have gas-to-core mass ratios that decrease with distance to values less than a percent.

As mentioned at the beginning of this subsection, the final state of an atmosphere that cools
and concomitantly accretes is an isothermal one. As such, isothermal simulations might
be expected to yield maximum gas-to-core mass ratios. If we interpret $\mu_{\rm iso}$
as shown in \figref{fig:mu} along these lines, then we might conclude that gas mass fractions of
super-Earth cores always remain much less than unity, even when such cores are embedded in
a gas-rich disk like the MMSN. This finding agrees with inferred
gas-to-core mass ratios of observed super-Earths \citep[e.g.,][]{Wu19}, but it would also imply that gas giants
cannot form at distances of a few au (where most are actually found; see
\citealt{Nielsen19}), 
unless the background disk were at least an order of magnitude more massive than the MMSN.
On the other hand, it also seems possible that $\mu_{\rm iso}$, though representing 
a maximally cooled state, does not necessarily equal the maximum $\mu$ possible. We return to this possibility in \secref{sec:accretion}.

\subsection{Flow Patterns}
\label{sec:flow}

\begin{figure*}
\includegraphics[width=1.99\columnwidth]{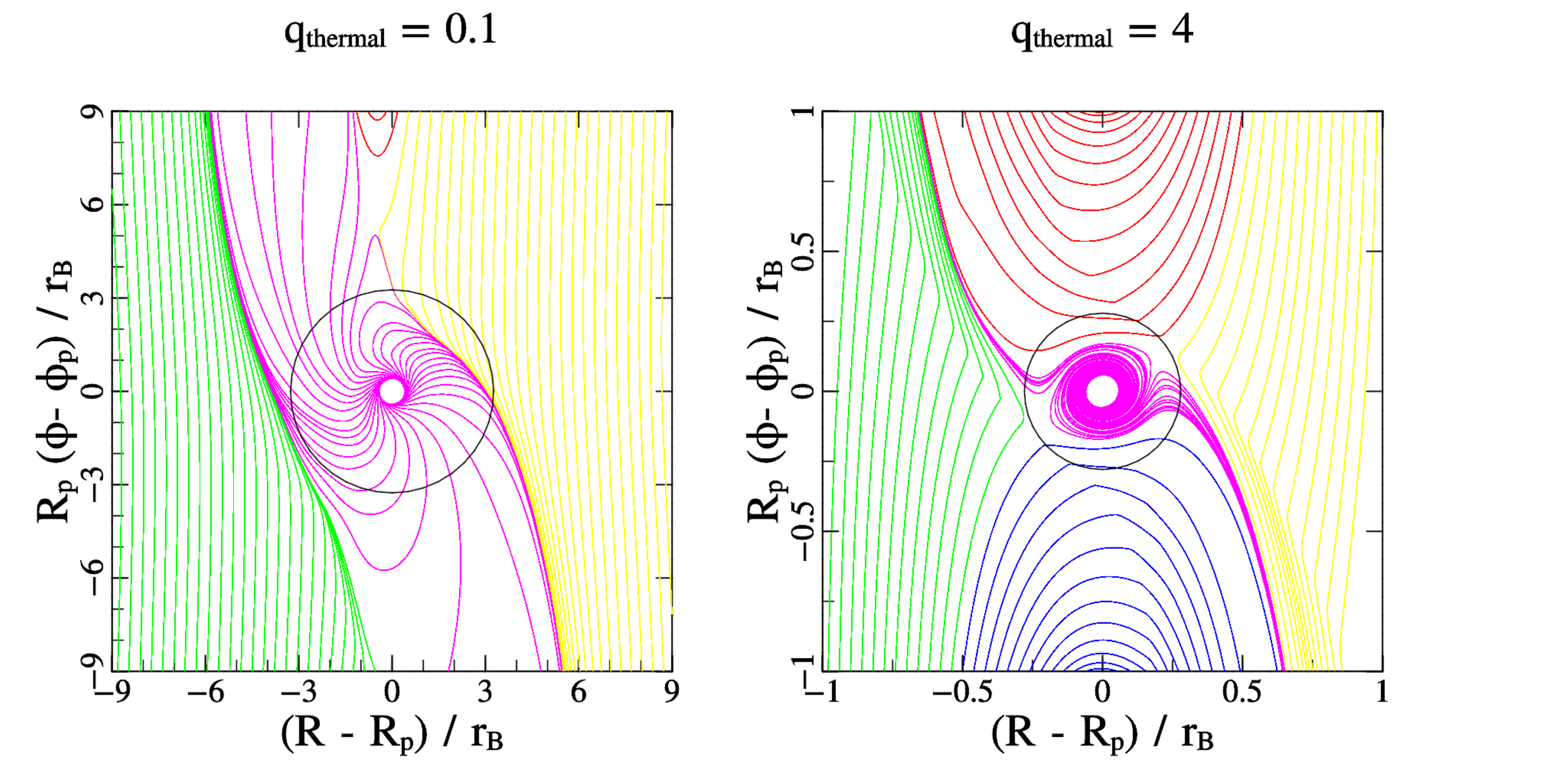}
\caption{Midplane streamlines in isothermal CPDs, showing models \#1 ($\qt=0.1$; left) and \#6 ($\qt=4$; right). The background Keplerian shear is from bottom to top in the inner disk ($R<\rp$), and top to bottom in the outer disk ($R>\rp$). The streamlines are color-coded: yellow and green are the inner and outer disk flow; red and blue are the inner and outer horseshoe flow; and magenta lines trace outflows away from the planet that are sourced from higher altitudes. Black circles mark the Hill radii of the planets. Gas exits the CPD near the L1 and L2 Lagrange points when the planet mass is superthermal, and these outflow paths widen as the planet mass decreases into the subthermal regime.}
\label{fig:stream_iso}
\end{figure*}

\begin{figure*}
\includegraphics[width=1.99\columnwidth]{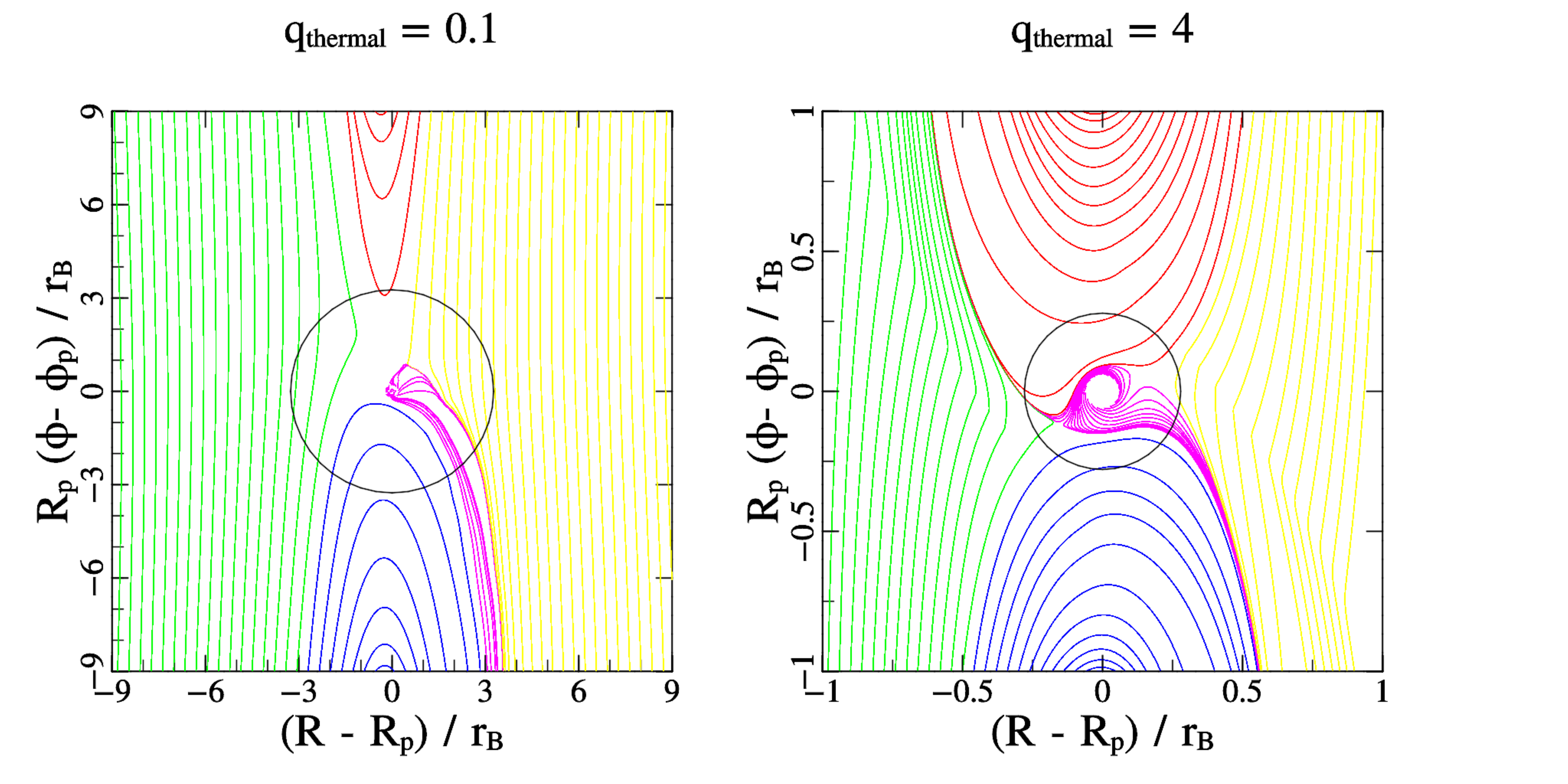}
\caption{Midplane streamlines in adiabatic CPDs, showing models \#7 ($\qt=0.1$; left) and \#12 ($\qt=4$; right). Streamlines are color-coded in the same manner as \figref{fig:stream_iso}. The outflow from the CPD (magenta lines) is mainly toward the outer disk. This is due to an asymmetry in the co-orbital dynamics, caused by the background entropy gradient \citep[e.g.][]{Paardekooper08,Masset09,Jimenez17}. Comparing the right panel here to the right panel of \figref{fig:stream_iso}, we find the two to be qualitatively similar, which suggests the EOS may be less important when the planet is superthermal.}
\label{fig:stream_ad}
\end{figure*}

\begin{figure*}
\includegraphics[width=1.99\columnwidth]{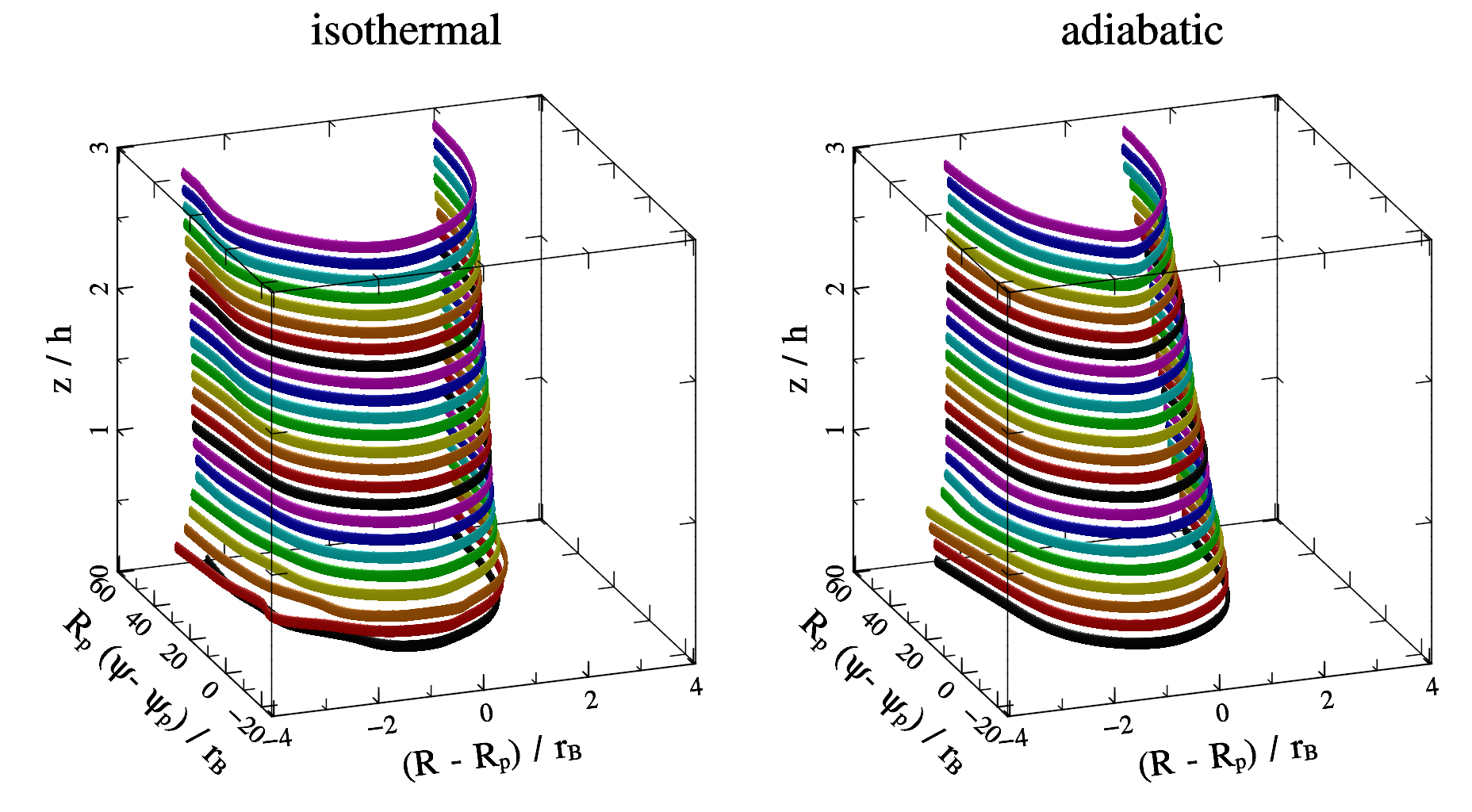}
\caption{Streamlines of the widest horseshoe orbits at different heights taken from models \#2 and \#8 where $\qt=0.25$. The flow is upstream where $R>\rp$ and downstream where $R<\rp$. On the left, we confirm that isothermal flow produces a columnar structure. On the right, we find that adiabtic flow is similarly columnar, but the width of horseshoe region gradually shrinks with height. At 3 scale heights, it shrinks to about half its width in the midplane.}
\label{fig:hs}
\end{figure*}

The flow structure in a CPD is close to axisymmetric within $\sim0.2\rB$, but becomes asymmetric as it merges with the background Keplerian flow. 
Figures \ref{fig:stream_iso} and \ref{fig:stream_ad} illustrate the flow patterns in the midplane for isothermal and adiabatic runs, respectively.
For isothermal gas, the outflow (traced by magenta lines) is directed predominantly through the L1 and L2 Lagrange points when $\qt > 1$ (\figref{fig:stream_iso} right panel); by comparison, when $\qt < 1$, these channels widen (\figref{fig:stream_iso} left panel).
This is expected because outflow speeds are generally subsonic, so if the gravitational potential at $\rH$ is much larger than the internal energy of the gas (so-called ``cold'' flows), then matter can only exit the Hill sphere near the Lagrange points. If instead the internal energy dominates, then it becomes possible to overflow the Hill sphere in all directions. This criterion to open up the outflow channel can be written as:
\begin{equation}
\frac{1}{c_{\rm iso}^2}\frac{GM}{\rH} = 3^{\frac{1}{3}} \qt^{\frac{2}{3}} > 1 \, .
\end{equation}
This translates to $\qt>0.6$, which we find to be consistent with our results.

The story is similar with the adiabatic cases, although there are some differences.
In both panels of \figref{fig:stream_ad}, we find the outflow to be focused toward the outer disk. 
The fact that this outflow connects to horseshoe orbits \citep[][]{Fung15} implies the outward horseshoe turns (blue streamlines that turn radially outward) are wider than the inward turns (red streamlines that turn radially inward). 
This asymmetry, which is strongest for subthermal
planets, has been seen in previous studies \citep[e.g.][]{Paardekooper08,Masset09,Jimenez17}, and is related to the entropy gradient in the PPD. 
Our setup introduces a positive radial entropy gradient when the gas is adiabatic, which is indeed expected to widen the outward horseshoe turns.

Interestingly, as planet mass increases and becomes superthermal, isothermal and adiabatic results seem to converge, as seen in the right panels of Figures \ref{fig:stream_iso} and \ref{fig:stream_ad}. This implies that when the planet is superthermal, CPD dynamics is dictated by gravity and the EOS is relegated to a more minor role.

We also look into vertical variations in the horseshoe orbits. \figref{fig:hs} plots the streamlines of the widest horseshoe orbits at different altitudes. \citet[][]{Fung15} and \citet{Masset16} showed that for isothermal disks, horseshoe orbits should align into columns. We confirm that this remains true in our isothermal simulations, as shown in the left panel of \figref{fig:hs}. \citet[][]{Fung15} suggested that it is an effect similar to Taylor--Proudman columns, and therefore might not apply to non-isothermal disks, where baroclinicity can alter the vorticity of the gas. The right panel of \figref{fig:hs} shows our results for an adiabatic case. The width of the horseshoe column gradually shrinks as altitude increases, and becomes about half its midplane value at 3 scale heights. We conlcude that while baroclinicity does introduce some variations, horseshoe orbits are still mostly columnar in adiabatic disks.

\subsection{Effects of Gap Opening}
\label{sec:gap}

\begin{figure*}
\includegraphics[width=1.99\columnwidth]{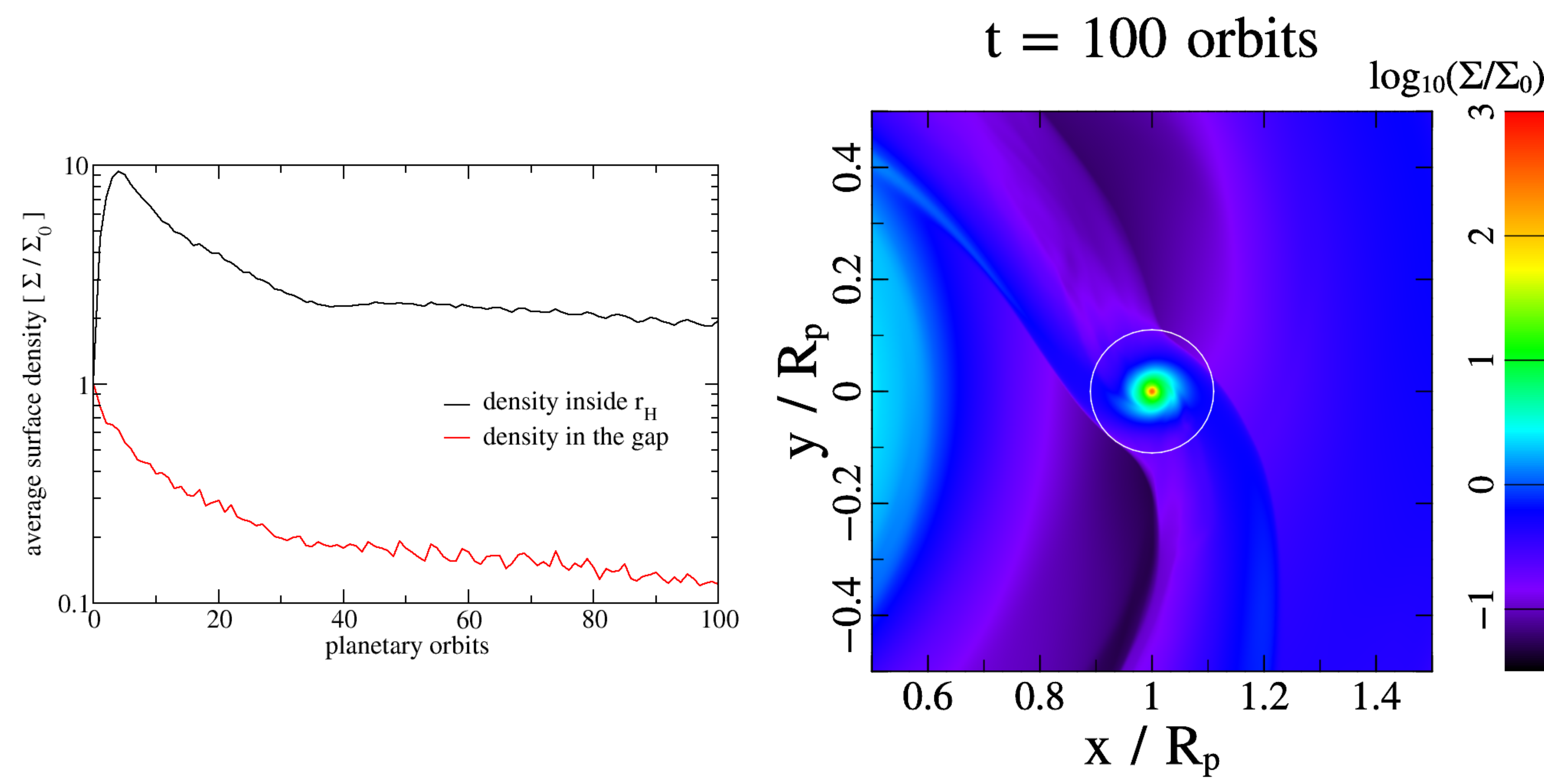}
\caption{Effects of gap-opening on the CPD, demonstrated using model \#14. On the left, we plot the average surface density inside the planet's Hill radius $\rH$ in black, and the average density in the PPD gap in red, both as functions of time. On the right is a 2D snapshot of the planet and the gap at 100 orbits, with the $\rH$ marked as the white circle. The ratio between the black and red lines remains constant over the majority of our simulation, despite the gap emptying by a factor of 10; the CPD appears well-coupled
to the background disk.}
\label{fig:gap}
\end{figure*}

Planets torque the gas in their orbits and open gaps in the PPD. In this section, we look into how gap opening affects the CPD. We have seen that the CPD mass---or equivalently, the gas-to-core mass ratio $\mu$---is proportional to the background density, as described by Equations \ref{eq:mu_iso} and \ref{eq:mu_ad}. From that, we can naively expect the CPD mass to decrease as the gap forms. To test this, we extend model \#14, which has the highest planet mass and exerts the strongest planetary torque, to 100 orbits.

In the left panel of \figref{fig:gap}, we compare how the mean surface densities of the CPD and of the PPD gap evolve with time. For the CPD, we compute the surface density by integrating the total mass within a
cylinder of radius of $1\,\rH$ around the planet and a vertical length equal to our simulation domain, and divide that by the surface area $2\pi\rH^2$. For the PPD gap density, we do the same for the region between $R = \{\rp-\rH,\,\rp+\rH\}$, $\Psi = \{\Psi_{\rm p}-0.5,\Psi_{\rm p}+0.5\}$, and $\theta = \{\pi/2-0.3, \pi/2\}$, 
with the CPD region excised.

The two surface densities evolve over time following a similar pattern. Since the sound crossing time in the CPD, 
$\sim\rH/c_{\rm iso}$, is about 1 $\Omega_{\rm p}^{-1}$, it is not surprising that the CPD reacts quickly to the emptying gap. We therefore conclude that \eqnref{eq:mu_iso} and \eqnref{eq:mu_ad} can also be used for gap-opening planets, as long as $\rho_0$ accounts for gap depletion.

Observationally, we know that PPDs dissipate over a few million years. Taken at face
value, our results imply that as PPDs
dissipate, CPDs should dissipate with them. In reality, however, we do not expect CPD evolution to play out so simply---in part because we have neglected cooling of the bound gas, which enables them to contract and survive the loss of external pressure from the dissipating nebula.
In 1D cooling models, the evolution of the atmosphere is
controlled by its radiative--convective boundary, whose
properties are insensitive to the nebular density at large 
\citep[e.g.,][]{Lee15}. Massive planetary envelopes
can accrete and survive even in nearly gas-free disks,
at least in 1D 
\citep[][]{Lee18}.
We will return to this tension between 1D cooling models
and 3D hydrodynamic models in Section \ref{sec:accretion}.

We note that our exploration of how PPD gaps influence CPDs is
also limited because our
simulations are optimized for smaller-scale CPDs and not for
larger-scale phenomena.
Our spatial resolution is poor far from the planet, with attendant
problems in numerical diffusion.
Moreover, 100 orbits is far from sufficient to 
evolve the gap to a steady state.

\section{Summary and Discussion}
\label{sec:conclude}
We have performed 3D hydrodynamics simulations of adiabatic and isothermal CPDs and demonstrated how their properties depend on 
$\qt$. We have also performed detailed resolution studies and compared data from three different codes, \texttt{PEnGUIn}, \texttt{Athena++}, and \texttt{Antares}.
We analyzed these results and established 
a general
understanding of CPD sizes, masses, and kinematics. To summarize:

\begin{itemize}
  \item Adiabatic CPDs are roughly spherically symmetric and bound within $\sim0.2$ $\rB$. Isothermal CPDs are bound within $\sim0.1$ $\rB$ and are rotationally supported inside $\sim0.05$ $\rB$. These scalings apply to subthermal ($\qt\leq1$) planets. Superthermal CPDs are smaller than these scalings predict.
  \item Rotational velocities
  in adiabatic CPDs scale linearly with $\qt$. If we extrapolate our results, adiabatic CPDs may become fully rotationally supported when $\qt\sim10$.
  \item The gas-to-core mass ratio, $\mu$, scales as $\qt^2$ when $\qt\lesssim1$, and $\qt^1$ when $\qt\gtrsim1$. Isothermal $\mu_{\rm iso}$'s are about 10 to 100 times higher than adiabatic $\mu_{\rm ad}$'s, but many orders of magnitude below what they would be if the isothermal CPDs were spherically symmetric and hydrostatic.
  \item In a minimum-mass solar nebula, $\mu_{\rm iso}$ is a few percent for cores of $\sim$10 $\Me$ near 1 au.
  \item Meridional flows around isothermal CPDs reach speeds of $4\sim5$ times the sound speed, while the flow speed around adiabatic CPDs is always subsonic.
  \item Gap opening does not decouple the CPD from the PPD, and so the CPD density remains proportional to the ambient gap density.
\end{itemize}

From a technical standpoint, we have also established that in order to fully capture CPD dynamics, simulations have to resolve scales as small as $0.05$ $\rB$. This is 
an expensive
requirement
in 3D; compared to resolving only $\rB$ (the typically assumed---and as we have shown, overestimated---CPD size for subthermal planets), the computational cost is $\sim20^4$
times higher. It is thanks to the advancement of computing technology that we are now capable of performing these simulations. 

Another equally important numerical parameter is $\rs$; whether CPDs are rotationally supported depends sensitively on its value. Typical values of $\rs$ used in the past have been around a few percent of min$(\rB,\,\rH)$ \citep[e.g.,][]{Fung15,Ormel15,Fung17a,Cimerman17,Lambrechts17,Lambrechts19}, which is large enough to erase rotationally supported disks. Physically, this means planets with core sizes larger than $0.05$ $\rB$ are unlikely to have rotationally supported disks. More tests with boundary conditions mimicking the core would be welcome (e.g., \citealt{Bethune19}).

Below, we discuss the implications of our results on gas giant formation, and compare our simulated CPDs to existing satellite systems.

\subsection{Forming Gas Giants}
\label{sec:accretion}
\begin{figure}
\includegraphics[width=0.99\columnwidth]{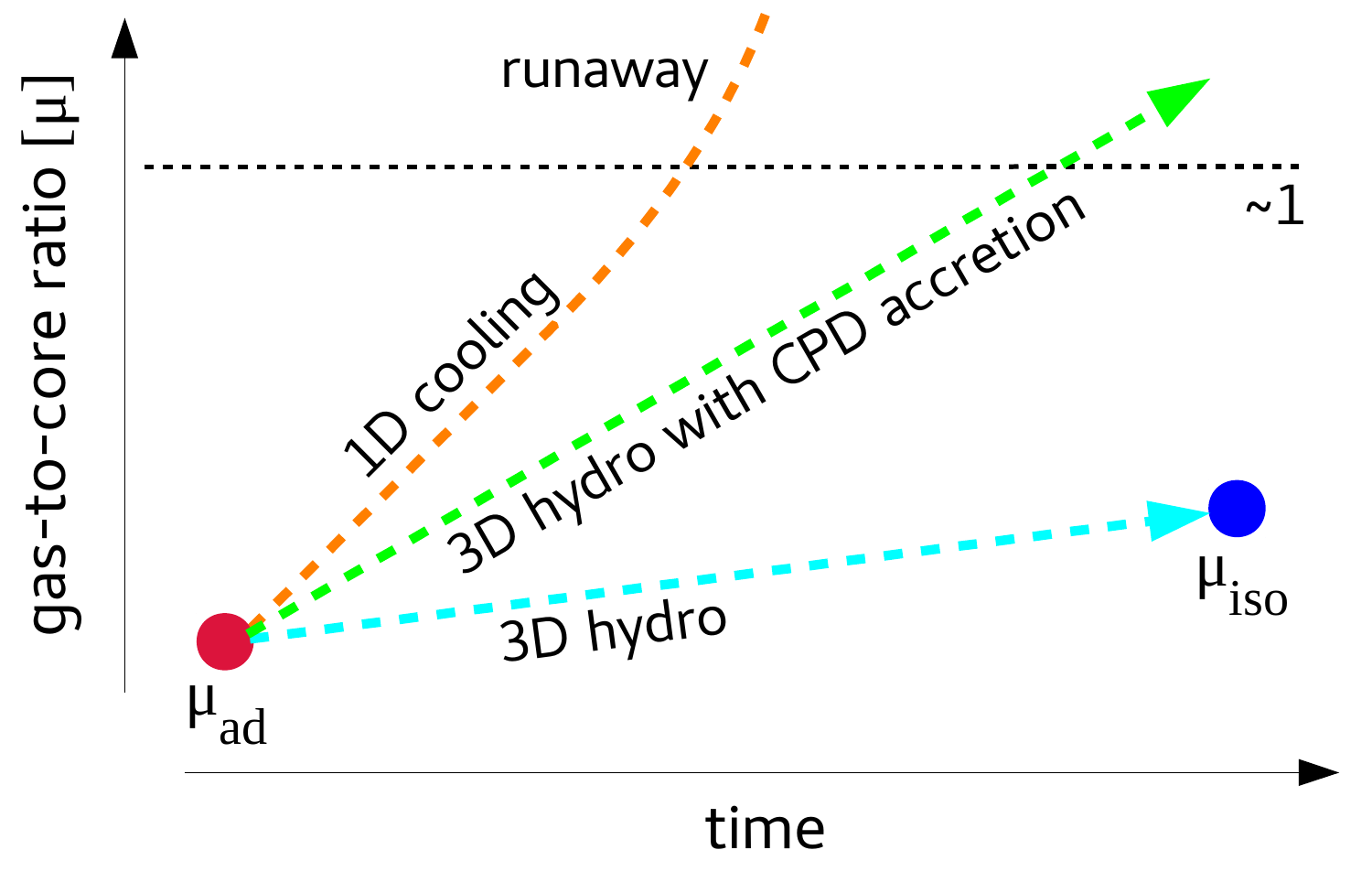} 
\caption{Schematic drawing of potential evolutionary paths for the gas-to-core mass ratio $\mu$. Planets start with $\mu_{\rm ad}$, and $\mu$ increases as the atmosphere cools and accretes. In classical 1D models, accretion is regulated by cooling, and, given enough time, can accumulate enough mass to enter the runaway regime, when $\mu \gtrsim 1$ (orange path). On the other hand, if our 3D $\mu_{\rm iso}$ corresponds to the final, maximally cooled state, then $\mu$ would instead evolve along the cyan path. The difference between the orange (1D) and cyan (3D hydro) paths is a consequence of 3D hydrodynamics. In principle, however, disk accretion physics within the CPD can boost $\mu_{\rm iso}$, perhaps to values crossing unity (green path).}
\label{fig:mu_draw}
\end{figure}

We found that the gas-to-core mass ratio $\mu$ remains below $10\%$ even for a 20 $\Me$ core surrounded by fully cooled, isothermal gas (right panel of \figref{fig:mu}). Theoretically, the adiabatic and isothermal cases should bracket a planet's thermal (read: accretion) history --- the atmosphere/CPD 
starts off behaving adiabatically on timescales shorter than the cooling time, and on timescales longer than the cooling time, becomes isothermal \citep[e.g.,][]{Lee14,Lee15,Ginzburg16,Coleman17}. Since $\mu$ is expected to evolve from the adiabatic to the isothermal state monotonically,\footnote{We have empirical evidence for monotonic evolution insofar as our experiments with perturbative cooling (not shown here) have yielded results intermediate between our adiabatic and isothermal runs.} then given our
result that $\mu_{\rm iso} < 10\%$, it would seem unlikely that envelope self-gravity would ever become significant enough to trigger ``runaway accretion'' and gas giant formation 
\citep[e.g.,][]{Pollack96,Ikoma00}.

Are there ways out of this conclusion? Is it possible
for $\mu_{\rm iso}$ to be larger than 10\%?
In 3D, we have seen that isothermal CPDs are rotationally supported within $\sim0.05\rB$. Rotationally supported envelopes can have arbitrary masses and density profiles (within the bounds of gravitational and hydrodynamic, e.g., Rayleigh stability).
Indeed, the densities given by \texttt{PEnGUIn} and \texttt{Athena++} do not agree inside
$0.05\rB$ (\figref{fig:iso_den_1D}). This leaves much room for speculation. An isothermal CPD could potentially become more massive if there are 
angular momentum transport mechanisms that allow it to accrete. \citet{Zhu16}, for example, reported shock-driven and vortex-driven accretion in their 2D simulations. Reality might be a mixture of the 1D and 3D models. The outer parts of the atmosphere (still within $0.05\rB$) may become radiative, nearly isothermal, and disky, with complex 3D flow structures like what we have seen in this work, while the inner parts may be convective, nearly adiabatic, and spherically symmetric. Gas might accrete across the isothermal disk and pile on top of the adiabatic envelope, in a fashion similar to 
the way circumstellar disks feed protostars.

We illustrate these ideas in \figref{fig:mu_draw} by drawing some schematic evolutionary paths. 1D models predict a cooling phase followed by a runaway phase after $\mu$ reaches unity (orange path).  Our 3D simulations, taken at face value, indicate that cooling alone leads to much smaller values of $\mu$ (cyan path). However, if one combines cooling, 3D hydrodynamics, and disk accretion physics, then one 
might produce an evolutionary path resembling the green path. Gas giants may 
form if disk accretion and eventually self-gravity push $\mu$ above unity.

\subsection{Comparisons with satellite systems}
\label{sec:satellite}
The presence of prograde, low-inclination, low-eccentricity ``regular'' satellites around the giant planets in our solar system suggests that there once existed rotationally supported CPDs around them, much like the ones we discover in our isothermal simulations. \citet{Canup06} found that the total mass of each satellite system 
is lower than its host's mass by a factor of $\sim10^{-4}$, which implies, if one assumes a gas-to-solid mass ratio of 100, a gaseous-CPD-to-planet mass ratio of $\sim10^{-2}$. This is encouragingly of the same order as our measured values for
$\mu_{\rm iso}$.

If satellite systems are formed in CPDs like the ones we simulated, we expect the former to have sizes comparable to or smaller than (if inward migration of solids is significant) 
$0.05~\rB$ (for $q_{\rm thermal} < 1$). We 
test this expectation here.
To evaluate $\rB$, one needs to estimate $c_{\rm iso}$ where the planets 
formed. For simplicity, we will assume they formed near their current positions, and use the same temperature profile as the one used in \secref{sec:mass}, where the temperature is about 300 K at $R = 1$ au  and scales as $R^{-1/2}$. We approximate the sizes of the regular satellite systems using the semi-major axes of their outermost members.

\subsubsection{Jupiter}
\label{sec:Jupiter}
The outermost prograde satellite orbiting Jupiter is Valetudo. Its orbital semi-major axis is $\sim1.9\times 10^7$ km \citep{Sheppard18}. At 5.2 au, $c_{\rm iso}$ in the original solar protoplanetary disk is approximately 660$\rm~m~s^{-1}$, which translates to $0.05~\rB \approx 1.5\times 10^7$ km. We note that Jupiter would be superthermal in our disk model, with $\qt=8$, which implies the CPD size could be smaller than $0.05\rB$.

\subsubsection{Saturn}
\label{sec:Saturn}
The outermost prograde satellite orbiting Saturn is Iapetus, with an orbital semi-major axis of $\sim3.6\times 10^6$ km \citep{Jacobson10}. At 9.5 au, we get $c_{\rm iso}\sim 570 {\rm~m~s^{-1}}$, which translates to $0.05~\rB \approx 5.8\times 10^6$ km. Saturn would have $\qt\sim1$ in our disk model. 

\subsubsection{Uranus}
\label{sec:Uranus}
The outermost prograde satellite orbiting Uranus is Oberon, with an orbital semi-major axis of $\sim5.8\times 10^5$ km \citep{Laskar87}. At 19.2 au, we get $c_{\rm iso}\sim 480 {\rm~m~s^{-1}}$, which translates to $0.05~\rB \approx 1.3\times 10^6$ km. Uranus would have $\qt\sim0.1$ in this model. 

\subsubsection{Neptune}
\label{sec:Neptune}
The outermost prograde satellite orbiting Neptune is Proteus, with an orbital semi-major axis of $\sim1.2\times 10^5$ km \citep{Jacobson04}. At 30.1 au, we get $c_{\rm iso}\sim 430 {\rm~m~s^{-1}}$, which translates to $0.05~\rB \approx 1.8\times 10^6$ km. Neptune would have $\qt\sim0.09$ in this model. 
Although our model disk is 15$\times$ larger than Neptune's actual satellite system, a complication arises from Triton, which 
lies just beyond Proteus at $\sim3.5\times 10^5$ km and
is suggested to be a captured satellite \citep{Agnor06}. It seems possible that Neptune once had a larger prograde satellite system, which was truncated when Triton was captured.

In summary, our estimated disk sizes are within a factor of 2 of the sizes of the prograde satellite systems around Jupiter, Saturn, and Uranus, and larger than Neptune's by an order of magnitude. This is consistent with these satellites having formed in CPDs like those in our isothermal simulations.


\acknowledgements
The authors thank Bertram Bitsch, Nicolas Cimerman, Sivan Ginzburg, Michiel Lambrechts, Chris Ormel, Tobias Moldenhauer and Yanqin Wu for encouraging discussions. We also thank Chun-Fan Liu and Hsien Shang for sharing simulation data. This work was partly performed under contract with the Jet Propulsion Laboratory (JPL) funded by NASA through the Sagan Fellowship Program executed by the NASA Exoplanet Science Institute. It was also partly performed at the Aspen Center for Physics, which is supported by National Science Foundation grant PHY-1607611.

\software{PEnGUIn \citep{MyThesis}, Athena++ code \citep{stone2008}, Antares \citep{Yuan05}}

\bibliographystyle{aasjournal}
\bibliography{Lit}

\begin{thebibliography}{}
\expandafter\ifx\csname natexlab\endcsname\relax\def\natexlab#1{#1}\fi

\bibitem[{{Agnor} \& {Hamilton}(2006)}]{Agnor06}
{Agnor}, C.~B., \& {Hamilton}, D.~P. 2006, \nat, 441, 192

\bibitem[{{Bate} {et~al.}(2003){Bate}, {Lubow}, {Ogilvie}, \&
  {Miller}}]{Bate03}
{Bate}, M.~R., {Lubow}, S.~H., {Ogilvie}, G.~I., \& {Miller}, K.~A. 2003,
  \mnras, 341, 213

\bibitem[{{B{\'e}thune} \& {Rafikov}(2019)}]{Bethune19}
{B{\'e}thune}, W., \& {Rafikov}, R.~R. 2019, \mnras, 488, 2365

\bibitem[{{Canup} \& {Ward}(2006)}]{Canup06}
{Canup}, R.~M., \& {Ward}, W.~R. 2006, \nat, 441, 834

\bibitem[{{Christiaens} {et~al.}(2019){Christiaens}, {Cantalloube}, {Casassus},
  {Price}, {Absil}, {Pinte}, {Girard}, \& {Montesinos}}]{Christiaens19}
{Christiaens}, V., {Cantalloube}, F., {Casassus}, S., {et~al.} 2019, \apjl,
  877, L33

\bibitem[{{Cimerman} {et~al.}(2017){Cimerman}, {Kuiper}, \&
  {Ormel}}]{Cimerman17}
{Cimerman}, N.~P., {Kuiper}, R., \& {Ormel}, C.~W. 2017, \mnras, 471, 4662

\bibitem[{{Coleman} {et~al.}(2017){Coleman}, {Papaloizou}, \&
  {Nelson}}]{Coleman17}
{Coleman}, G. A.~L., {Papaloizou}, J. C.~B., \& {Nelson}, R.~P. 2017, \mnras,
  470, 3206

\bibitem[{{D'Angelo} \& {Bodenheimer}(2013)}]{DAngelo13}
{D'Angelo}, G., \& {Bodenheimer}, P. 2013, \apj, 778, 77

\bibitem[{{D'Angelo} {et~al.}(2003){D'Angelo}, {Kley}, \&
  {Henning}}]{DAngelo03}
{D'Angelo}, G., {Kley}, W., \& {Henning}, T. 2003, \apj, 586, 540

\bibitem[{{Fung}(2015)}]{MyThesis}
{Fung}, J. 2015, PhD thesis, University of Toronto, Canada

\bibitem[{{Fung} {et~al.}(2015){Fung}, {Artymowicz}, \& {Wu}}]{Fung15}
{Fung}, J., {Artymowicz}, P., \& {Wu}, Y. 2015, \apj, 811, 101

\bibitem[{{Fung} {et~al.}(2017){Fung}, {Masset}, {Lega}, \&
  {Velasco}}]{Fung17a}
{Fung}, J., {Masset}, F., {Lega}, E., \& {Velasco}, D. 2017, \aj, 153, 124

\bibitem[{{Ginzburg} {et~al.}(2016){Ginzburg}, {Schlichting}, \&
  {Sari}}]{Ginzburg16}
{Ginzburg}, S., {Schlichting}, H.~E., \& {Sari}, R. 2016, \apj, 825, 29

\bibitem[{{Haffert} {et~al.}(2019){Haffert}, {Bohn}, {de Boer}, {Snellen},
  {Brinchmann}, {Girard}, {Keller}, \& {Bacon}}]{Haffert19}
{Haffert}, S.~Y., {Bohn}, A.~J., {de Boer}, J., {et~al.} 2019, Nature
  Astronomy, 329

\bibitem[{{Ikoma} {et~al.}(2000){Ikoma}, {Nakazawa}, \& {Emori}}]{Ikoma00}
{Ikoma}, M., {Nakazawa}, K., \& {Emori}, H. 2000, \apj, 537, 1013

\bibitem[{{Jacobson}(2010)}]{Jacobson10}
{Jacobson}, R. 2010, JPL satellite ephemeris, SAT339

\bibitem[{{Jacobson} \& {Owen}(2004)}]{Jacobson04}
{Jacobson}, R.~A., \& {Owen}, Jr., W.~M. 2004, \aj, 128, 1412

\bibitem[{{Jim{\'e}nez} \& {Masset}(2017)}]{Jimenez17}
{Jim{\'e}nez}, M.~A., \& {Masset}, F.~S. 2017, \mnras, 471, 4917

\bibitem[{{Keppler} {et~al.}(2018){Keppler}, {Benisty}, {M{\"u}ller},
  {Henning}, {van Boekel}, {Cantalloube}, {Ginski}, {van Holstein}, {Maire},
  {Pohl}, {Samland }, {Avenhaus}, {Baudino}, {Boccaletti}, {de Boer},
  {Bonnefoy}, {Chauvin}, {Desidera}, {Langlois}, {Lazzoni}, {Marleau},
  {Mordasini}, {Pawellek}, {Stolker}, {Vigan}, {Zurlo}, {Birnstiel},
  {Brandner}, {Feldt}, {Flock}, {Girard}, {Gratton}, {Hagelberg}, {Isella},
  {Janson}, {Juhasz}, {Kemmer}, {Kral}, {Lagrange}, {Launhardt}, {Matter},
  {M{\'e}nard}, {Milli}, {Molli{\`e}re}, {Olofsson}, {P{\'e}rez}, {Pinilla},
  {Pinte}, {Quanz}, {Schmidt}, {Udry}, {Wahhaj}, {Williams}, {Buenzli},
  {Cudel}, {Dominik}, {Galicher}, {Kasper}, {Lannier}, {Mesa}, {Mouillet},
  {Peretti}, {Perrot}, {Salter}, {Sissa}, {Wildi}, {Abe}, {Antichi},
  {Augereau}, {Baruffolo}, {Baudoz}, {Bazzon}, {Beuzit}, {Blanchard}, {Brems},
  {Buey}, {De Caprio}, {Carbillet}, {Carle}, {Cascone}, {Cheetham}, {Claudi},
  {Costille}, {Delboulb{\'e}}, {Dohlen}, {Fantinel}, {Feautrier}, {Fusco},
  {Giro}, {Gluck}, {Gry}, {Hubin}, {Hugot}, {Jaquet}, {Le Mignant}, {Llored},
  {Madec}, {Magnard}, {Martinez}, {Maurel}, {Meyer}, {M{\"o}ller-Nilsson},
  {Moulin}, {Mugnier}, {Orign{\'e}}, {Pavlov}, {Perret}, {Petit}, {Pragt},
  {Puget}, {Rabou}, {Ramos}, {Rigal}, {Rochat}, {Roelfsema}, {Rousset}, {Roux},
  {Salasnich}, {Sauvage}, {Sevin}, {Soenke}, {Stadler}, {Suarez}, {Turatto}, \&
  {Weber}}]{Keppler18}
{Keppler}, M., {Benisty}, M., {M{\"u}ller}, A., {et~al.} 2018, \aap, 617, A44

\bibitem[{{Kley}(1998)}]{Kley98}
{Kley}, W. 1998, \aap, 338, L37

\bibitem[{{Kurokawa} \& {Tanigawa}(2018)}]{Kurokawa18}
{Kurokawa}, H., \& {Tanigawa}, T. 2018, \mnras, 479, 635

\bibitem[{{Kuwahara} {et~al.}(2019){Kuwahara}, {Kurokawa}, \&
  {Ida}}]{Kuwahara19}
{Kuwahara}, A., {Kurokawa}, H., \& {Ida}, S. 2019, \aap, 623, A179

\bibitem[{{Lambrechts} \& {Lega}(2017)}]{Lambrechts17}
{Lambrechts}, M., \& {Lega}, E. 2017, \aap, 606, A146

\bibitem[{{Lambrechts} {et~al.}(2019){Lambrechts}, {Lega}, {Nelson}, {Crida},
  \& {Morbidelli}}]{Lambrechts19}
{Lambrechts}, M., {Lega}, E., {Nelson}, R.~P., {Crida}, A., \& {Morbidelli}, A.
  2019, \aap, 630, A82

\bibitem[{{Laskar} \& {Jacobson}(1987)}]{Laskar87}
{Laskar}, J., \& {Jacobson}, R.~A. 1987, \aap, 188, 212

\bibitem[{{Lee} \& {Chiang}(2015)}]{Lee15}
{Lee}, E.~J., \& {Chiang}, E. 2015, \apj, 811, 41

\bibitem[{{Lee} {et~al.}(2018){Lee}, {Chiang}, \& {Ferguson}}]{Lee18}
{Lee}, E.~J., {Chiang}, E., \& {Ferguson}, J.~W. 2018, \mnras, 476, 2199

\bibitem[{{Lee} {et~al.}(2014){Lee}, {Chiang}, \& {Ormel}}]{Lee14}
{Lee}, E.~J., {Chiang}, E., \& {Ormel}, C.~W. 2014, \apj, 797, 95

\bibitem[{{Machida} {et~al.}(2008){Machida}, {Kokubo}, {Inutsuka}, \&
  {Matsumoto}}]{Machida08}
{Machida}, M.~N., {Kokubo}, E., {Inutsuka}, S.-i., \& {Matsumoto}, T. 2008,
  \apj, 685, 1220

\bibitem[{{Martin} \& {Lubow}(2011)}]{Martin11}
{Martin}, R.~G., \& {Lubow}, S.~H. 2011, \mnras, 413, 1447

\bibitem[{{Masset} \& {Ben{\'{\i}}tez-Llambay}(2016)}]{Masset16}
{Masset}, F.~S., \& {Ben{\'{\i}}tez-Llambay}, P. 2016, \apj, 817, 19

\bibitem[{{Masset} \& {Casoli}(2009)}]{Masset09}
{Masset}, F.~S., \& {Casoli}, J. 2009, \apj, 703, 857

\bibitem[{{Masset} {et~al.}(2006){Masset}, {D'Angelo}, \& {Kley}}]{Masset06}
{Masset}, F.~S., {D'Angelo}, G., \& {Kley}, W. 2006, \apj, 652, 730

\bibitem[{{Nielsen} {et~al.}(2019){Nielsen}, {De Rosa}, {Macintosh}, {Wang},
  {Ruffio}, {Chiang}, {Marley}, {Saumon}, {Savransky}, {Ammons}, {Bailey},
  {Barman}, {Blain}, {Bulger}, {Burrows}, {Chilcote}, {Cotten}, {Czekala},
  {Doyon}, {Duch{\^e}ne}, {Esposito}, {Fabrycky}, {Fitzgerald}, {Follette},
  {Fortney}, {Gerard}, {Goodsell}, {Graham}, {Greenbaum}, {Hibon}, {Hinkley},
  {Hirsch}, {Hom}, {Hung}, {Dawson}, {Ingraham}, {Kalas}, {Konopacky},
  {Larkin}, {Lee}, {Lin}, {Maire}, {Marchis}, {Marois}, {Metchev},
  {Millar-Blanchaer}, {Morzinski}, {Oppenheimer}, {Palmer}, {Patience},
  {Perrin}, {Poyneer}, {Pueyo}, {Rafikov}, {Rajan}, {Rameau}, {Rantakyr{\"o}},
  {Ren}, {Schneider}, {Sivaramakrishnan}, {Song}, {Soummer}, {Tallis},
  {Thomas}, {Ward-Duong}, \& {Wolff}}]{Nielsen19}
{Nielsen}, E.~L., {De Rosa}, R.~J., {Macintosh}, B., {et~al.} 2019, \aj, 158,
  13

\bibitem[{{Ormel} {et~al.}(2015){Ormel}, {Shi}, \& {Kuiper}}]{Ormel15}
{Ormel}, C.~W., {Shi}, J.-M., \& {Kuiper}, R. 2015, \mnras, 447, 3512

\bibitem[{{Paardekooper} \& {Mellema}(2008)}]{Paardekooper08}
{Paardekooper}, S.-J., \& {Mellema}, G. 2008, \aap, 478, 245

\bibitem[{{Pollack} {et~al.}(1996){Pollack}, {Hubickyj}, {Bodenheimer},
  {Lissauer}, {Podolak}, \& {Greenzweig}}]{Pollack96}
{Pollack}, J.~B., {Hubickyj}, O., {Bodenheimer}, P., {et~al.} 1996, \icarus,
  124, 62

\bibitem[{{Quillen} \& {Trilling}(1998)}]{Quillen98}
{Quillen}, A.~C., \& {Trilling}, D.~E. 1998, \apj, 508, 707

\bibitem[{{Schulik} {et~al.}(2019){Schulik}, {Johansen}, {Bitsch}, \&
  {Lega}}]{Schulik19}
{Schulik}, M., {Johansen}, A., {Bitsch}, B., \& {Lega}, E. 2019, arXiv
  e-prints, arXiv:1909.08359

\bibitem[{{Sheppard} {et~al.}(2018){Sheppard}, {Trujillo}, \&
  {Williams}}]{Sheppard18}
{Sheppard}, S.~S., {Trujillo}, C., \& {Williams}, G.~V. 2018, Minor Planet
  Electronic Circulars, 2018-O09

\bibitem[{{Stone} {et~al.}(2008){Stone}, {Gardiner}, {Teuben}, {Hawley}, \&
  {Simon}}]{stone2008}
{Stone}, J.~M., {Gardiner}, T.~A., {Teuben}, P., {Hawley}, J.~F., \& {Simon},
  J.~B. 2008, \apjs, 178, 137

\bibitem[{{Szul{\'a}gyi}(2017)}]{Szulagyi17}
{Szul{\'a}gyi}, J. 2017, \apj, 842, 103

\bibitem[{{Szul{\'a}gyi} {et~al.}(2016){Szul{\'a}gyi}, {Masset}, {Lega},
  {Crida}, {Morbidelli}, \& {Guillot}}]{Szulagyi16}
{Szul{\'a}gyi}, J., {Masset}, F., {Lega}, E., {et~al.} 2016, \mnras, 460, 2853

\bibitem[{{Tanigawa} {et~al.}(2012){Tanigawa}, {Ohtsuki}, \&
  {Machida}}]{Tanigawa12}
{Tanigawa}, T., {Ohtsuki}, K., \& {Machida}, M.~N. 2012, \apj, 747, 47

\bibitem[{{Wagner} {et~al.}(2018){Wagner}, {Follete}, {Close}, {Apai}, {Gibbs},
  {Keppler}, {M{\"u}ller}, {Henning}, {Kasper}, {Wu}, {Long}, {Males},
  {Morzinski}, \& {McClure}}]{Wagner18}
{Wagner}, K., {Follete}, K.~B., {Close}, L.~M., {et~al.} 2018, \apjl, 863, L8

\bibitem[{{Wang} {et~al.}(2014){Wang}, {Bu}, {Shang}, \& {Gu}}]{Wang14}
{Wang}, H.-H., {Bu}, D., {Shang}, H., \& {Gu}, P.-G. 2014, \apj, 790, 32

\bibitem[{{Wu}(2019)}]{Wu19}
{Wu}, Y. 2019, \apj, 874, 91

\bibitem[{{Yuan} \& {Yen}(2005)}]{Yuan05}
{Yuan}, C., \& {Yen}, D. C.~C. 2005, Journal of Korean Astronomical Society,
  38, 197

\bibitem[{{Zhang} {et~al.}(2018){Zhang}, {Zhu}, {Huang}, {Guzm{\'a}n},
  {Andrews}, {Birnstiel}, {Dullemond}, {Carpenter}, {Isella}, {P{\'e}rez},
  {Benisty}, {Wilner}, {Baruteau}, {Bai}, \& {Ricci}}]{Zhang18}
{Zhang}, S., {Zhu}, Z., {Huang}, J., {et~al.} 2018, \apjl, 869, L47

\bibitem[{{Zhu} {et~al.}(2016){Zhu}, {Ju}, \& {Stone}}]{Zhu16}
{Zhu}, Z., {Ju}, W., \& {Stone}, J.~M. 2016, \apj, 832, 193

\bibitem[{{Zhu} \& {Stone}(2018)}]{zhu2018}
{Zhu}, Z., \& {Stone}, J.~M. 2018, \apj, 857, 34

\end{thebibliography}

\end{CJK*}
\end{document}